\documentclass[aps,pra,twocolumn,10pt,showpacs,amsmath,amssymb,floatfix,superscriptaddress,longbibliography]{revtex4-1}
\usepackage{graphicx}
\usepackage{amsmath}
\usepackage{amssymb}
\usepackage[utf8]{inputenc}
\usepackage{comment}
\usepackage{dcolumn}
\usepackage{bm}
\usepackage{hyperref}
\usepackage{color}
\usepackage{soul}

\newcommand{\Rmnum}[1]{\expandafter\@slowromancap\romannumeral  #1@}

\newcommand{\xc}{\textrm{xc}}
\newcommand{\Hxc}{\textrm{Hxc}}
\newcommand{\br}{\mathbf{r}}
\newcommand{\bs}{\mathbf{s}}
\newcommand{\bn}{\mathbf{n}}
\newcommand{\bmm}{\mathbf{m}}

\newcommand{\Eq}[1]{Eq.(\ref{#1})}

\usepackage{bbold}
\newcommand{\rb}{\mathbf{r}}
\newcommand{\frm}{\textrm{f}}
\newcommand{\ordo}[1]{\mathcal{O}\left ( #1 \right )}

\graphicspath{{Figures/}}
\begin{document}

\title{Time-dependent density-functional theory for strongly interacting electrons}

\author{Luis Cort} 
\affiliation{Department of Physics,
Nanoscience Center P.O.Box 35 FI-40014 University of Jyv\"{a}skyl\"{a}, Finland}

\author{Daniel Karlsson} 
\affiliation{Department of Physics,
Nanoscience Center P.O.Box 35 FI-40014 University of Jyv\"{a}skyl\"{a}, Finland}

\author{Giovanna Lani} 
\affiliation{lnstitut de Min{\'e}ralogie, de  Physique des Mat{\'e}riaux et de Cosmochimie, Universit{\'e} Pierre et Marie Curie, 4 place Jussieu, 75252, Paris cedex 05, France}

\author{Robert van Leeuwen}
\affiliation{Department of Physics,
Nanoscience Center P.O.Box 35 FI-40014 University of Jyv\"{a}skyl\"{a}, Finland}

\begin{abstract}
We consider an analytically solvable model of two interacting electrons that allows for the calculation of the exact exchange-correlation kernel of time-dependent density functional theory.
This kernel, as well as the corresponding density response function, is studied in the limit of large repulsive interactions between the electrons and we give analytical
results for these quantities as an asymptotic expansion in powers of the square root of the interaction strength.
We find that in the strong interaction limit the three leading terms in the expansion of the kernel act instantaneously
while memory terms only appear in the next orders.
We further derive an alternative expansion for the kernel in the strong interaction limit on the basis of the theory developed in [Phys. Chem. Chem. Phys. {\bf 18}, 21092 (2016)] using 
the formalism of strictly correlated electrons in the adiabatic approximation.
We find that the first two leading terms in this series, corresponding to the strictly correlated limit and its zero-point vibration correction, coincide
with the two leading terms of the exact expansion. 
We finally analyze the spatial non-locality of these terms and show when the adiabatic approximation breaks down. 
The ability to reproduce the exact kernel in the strong interaction limit
indicates that the adiabatic strictly correlated electron formalism is useful for studying the density response and excitation 
properties of other systems with strong electronic interactions.

\end{abstract}

\maketitle

\section{\textbf{Introduction}}
Time-dependent density functional theory (TDDFT) \cite{Gross1996,Burke2005,Leeuwen2001,Ullrich2012,Marques2012,Ruggenthaler2015,Maitra2016} is a well-established approach to study the time-dependent and excitation properties of many-electron systems. One of the main reasons for its popularity is that within this formalism the time-dependent  interacting many-body problem can be recast exactly into an equivalent one-particle framework, advantageous for numerical implementations. The corresponding one-particle equations, called the time-dependent Kohn-Sham equations, contain an effective potential, known as the Kohn-Sham potential, which is defined in such a way that the non-interacting system has the same time-dependent density as the original interacting many-body system. The Kohn-Sham potential is typically written as the sum of the external potential of the interacting system and of the Hartree and the exchange-correlation (xc) potential.

In practical applications of TDDFT, the xc-potential $v_{\xc}$ is approximated. The type of approximation employed crucially determines the quality of the results, and therefore a considerable amount of research has gone into the difficult task of finding reliable and accurate approximations for this quantity. This task is simpler in the linear response regime where we consider small variations in the density caused by applied perturbations.  
This regime is of interest as the knowledge of the density response function is sufficient to calculate the excitation energies and the absorption spectrum of the system~\cite{Petersilka1996}. For this purpose it is enough to know $v_\xc$ and its functional derivative $\delta v_\xc/\delta n = f_\xc$ with respect to the density $n$, evaluated at the ground-state density. The quantity $f_\xc$ is called the xc-kernel and has been the subject of intense investigations.

The simplest possible approximation is the adiabatic local-density approximation (ALDA) in which the xc-kernel is local in space and time. The ALDA has, however, a number of deficiencies~\cite{Ullrich2012} such as, for example, the inability to produce correct charge transfer excitations \cite{Stein2009,Gritsenko2004,Maitra2005,Autschbach2009} and Born-Oppenheimer surfaces of excited states in dissociating molecules \cite{Gritsenko2000,Hieringer2006,Giesbertz2008} and semiconductor band gaps~\cite{Botti2004}. Some improvements have been made using hybrid functionals which contain mixtures of exact exchange and traditional local functionals. These methods are non-local in space but still adiabatic. However, they are not systematic and the optimal mixture of exact-exchange is often system-dependent~\cite{Dreuw2003,Hofmann2012a,Dierksen2004,Peach2008}. Other, more systematic, approximations for $f_\xc$ beyond the ALDA often rely on perturbative expansions \cite{Gorling1998,Tokatly2001,VonBarth2005,Fuchs2005,Hellgren2009,Hellgren2010,Hesselmann2011,Hesselmann2011b,Bates2013,Erhard2016} and many of them are restricted to the exchange-only approximation.
Their perturbative nature makes these approaches questionable
in the strong correlation regime which is relevant in various physical situations, notably the case of molecular dissociation, and hence it is highly desirable to develop new techniques to tackle this regime. 

In a recent work~\cite{Lani2016} the so-called strictly correlated electrons (SCE) framework \cite{Seidl1999,Gori-Giorgi2009b,Malet2012a}, a formalism well suited for the description of strong interactions, has been applied within the time-dependent domain. The authors derived an expression for the xc kernel in the so-called adiabatic approximation 
and established that the adiabatic SCE (ASCE) kernel satisfies the zero force theorem \cite{Mundt2007}, an exact property related to generalized translational invariance \cite{Vignale1995a,Dobson1994}. 
The kernel was furthermore studied for finite one-dimensional systems with different density profiles, some of which are prototypical of the dissociation of two-electron homonuclear molecules. It was found that the ASCE kernel is spatially non-local and exhibits a divergent behavior as the molecular bond is stretched. For adiabatic kernels this diverging behavior is crucial~\cite{Gritsenko2000} for describing bond-breaking excitations, which is a notoriously challenging problem in linear response TDDFT.
Since the kernel was derived in the adiabatic approximation, not much could be concluded about the limitations of its adiabatic nature in the context of the physics of strong static correlation. 
The case of infinitely strong electron-electron correlation is quite peculiar
and to date it is not known how accurate the adiabatic approximation can be in such a regime.

One way to shed light on this issue is to benchmark the ASCE kernel against an exact expression for $f_{\xc}$ obtained from a model system where the density response function and thus the xc-kernel can be calculated analytically. In this work we consider such a model, namely two interacting electrons on a quantum ring \cite{Viefers2004,Ruggenthaler2012a,Loos2012,Ruggenthaler2015}, for which not only we compute the exact density response function and xc-kernel, but we also obtain these quantities for various two-body interaction strengths, including the infinitely strong one, and we compare these results with those given by the SCE theory in the adiabatic approximation. 
The leading order of the asymptotic expansion for the exact $f_\xc$ and the expression for the ASCE kernel are found to be identical. We also derive the next order correction term beyond the ASCE, called the 
adiabatic zero-point-energy approximation (AZPE), and show that also this adiabatic term is the same as the next order from the asymptotic expansion. The third order in the expansion for the kernel is 
still adiabatic, while a frequency dependence appears in the fourth order. In this order, an adiabatic approximation would break down.
This is one of the central results of the paper and elucidates both the strengths and the weaknesses of the adiabatic approximation in the limit of strong electron-electron interaction.

The paper is organized as follows. In Sec.~\ref{sec1} we introduce the quantum ring model and, after computing
the full spectrum of its Hamiltonian, we discuss asymptotic expansions for its eigenenergies and eigenstates in the case of strong interactions and analyze them. In Sec.~\ref{secDensityResponse} we study the density response function of strongly interacting systems, while in Sec.~\ref{secTDDFTStrong} we focus on TDDFT in the same regime and give an asymptotic expansion for the xc-kernel. Our conclusions are finally presented in Sec.~\ref{secConclusions}.

\section{\textbf{An exactly solvable system}}
\label{sec1}
\subsection{Two interacting electrons on a quantum ring}

For our study of electron correlations we consider an analytically solvable model, which we will refer to as the quantum ring model, of two electrons on a ring of length $L$ which repel each other with a two-body interaction.
The interaction strength can be adjusted using a parameter, which allows us to study the exact properties of the system ranging from
weak to very strong interactions. The explicit form of the Hamiltonian of the quantum ring is given by:
\begin{equation}
\hat{H}=-\frac{1}{2}(\partial_{x_{1}}^2+\partial_{x_{2}}^2)+\lambda V_0 \, \cos^2\left[\frac{\pi}{L}(x_{1}-x_{2})\right]   
\label{eq:Hamiltonian}
\end{equation}
where $\lambda \geq 0$ is a dimensionless parameter and $V_0$ has units of energy. The coordinates $x_1$ and $x_2$ are the coordinates of the electrons on the ring which run from $0$ to $L$. 
The ground-state density $n_0=2/L$ is spatially constant and independent of $\lambda$; for this reason the model can be used to illustrate several features of the coupling strength dependence in density functional theory. 

In order to calculate the properties of the system we have to determine the eigenfunctions $\Psi$ 
which satisfy the stationary Schr\"odinger equation $\hat{H} \Psi = E \Psi$ where $E$ are the energy eigenvalues.
For our two-particle system these eigenfunctions can be written as a product of a spatial wave function and a spin function as follows
\[
\Psi(x_{1}\sigma_1,x_{2}\sigma_2)=\psi^{\pm}(x_{1},x_{2}) \, \Xi^{\pm} (\sigma_1 , \sigma_2).
\]
For the singlet case (which we will focus on) the normalized spin function is given by
\[
\Xi^{+} (\sigma_1, \sigma_2) =\frac{1}{\sqrt{2}}\left(\delta_{\sigma_1\uparrow}\delta_{\sigma_2\downarrow} - \delta_{\sigma_1\downarrow}\delta_{\sigma_2\uparrow}\right)
\label{singlet}
\]
and is anti-symmetric in the spin variables.
For the triplet there are three linearly independent symmetric spin functions which we, for simplicity, all denote by $\Xi^{-} (\sigma_1,\sigma_2)$. 
Since the two-electron wave function $\Psi$ is anti-symmetric under the simultaneous interchange of space and spin variables it follows that
the spatial wave functions $\psi^\pm$ satisfy the symmetry relation
\[
\psi^\pm (x_1,x_2) = \pm \, \psi^\pm (x_2,x_1).
\]
Apart from these symmetry conditions, the Schr\"odinger equation needs to be solved with periodic boundary conditions on the
variables $x_1$ and $x_2$, i.e. the wave function and its first spatial derivatives are invariant under the substitution $x_i \rightarrow x_i+L$ for $i=1,2$. Using these conditions we can solve the Schr\"odinger equation by a suitable coordinate transformation. Since these steps are carried out in detail in Ref.~\cite{Ruggenthaler2013} here we just outline the main steps relevant for this work.

The Hamiltonian (\ref{eq:Hamiltonian}) becomes separable in the terms $R=(x_1+x_2)/2$, the
center-of-mass coordinate, and $z=\pi (x_1-x_2)/L$, the
dimensionless relative coordinate.
This variable transformation gives
\begin{equation}
\hat{H}=-\frac{1}{4}\partial_{R}^2- \frac{\pi^2}{L^2} \partial_{z}^2+\lambda V_0 \cos^2 (z) .  \nonumber
\label{eq:Hamiltonian2}
\end{equation}
By inserting a product Ansatz of the form $\psi (R,z) =  f(R) M (z)$ into the Schr\"odinger equation 
we find that the spatial two-particle eigenfunctions are
of the form
\begin{equation}
\psi (R,z) = \exp {\left(\frac{2\pi i k R}{L}\right)} \, M(z)
\label{eq:ansatz}
\end{equation}
where $k$ is an integer and the function $M(z)$ satisfies the Mathieu equation 
which we write in its standard form as \cite{Arscott1964, NIST}
\begin{equation}
\left[ - \partial_z^2   +  2 q \, \cos (2z) \right] M (z) =  a \, M(z) 
\label{eq:Mathieu}
\end{equation}
where the constants $q$ and $a$ are given by:
\begin{align}
q & = \lambda V_0 \left ( \frac{ L}{2\pi} \right )^2
\label{eq:q_def} \\
a & = - k^2  -2q + \frac{E L^2}{\pi^2} .
\label{eq:mathieu_characteristic}
\end{align}
For a given value of $q$ the Mathieu equation (\ref{eq:Mathieu}) has only periodic solutions for particular values 
$a(q)$ which are called the Mathieu characteristic values. Moreover this equation has either even or odd periodic solutions
which are called the Mathieu cosine and Mathieu sine functions respectively. Both sets of functions form a countable set, therefore
its members can be labeled by a non-negative integer $l$. For the Mathieu cosines this label starts at $l=0$ and
for the Mathieu sines at $l=1$.
The even Mathieu cosine function is denoted by $C_l (z;q)$ and its characteristic
value by $a_l^+ (q)$, while the odd Mathieu sine function is denoted by $S_l (z;q)$ and its characteristic value by $a^-_l (q)$. 
Since the center-of-mass wave functions are symmetric under the interchange of the spatial
coordinates of the electrons we see from Eq.(\ref{eq:ansatz}) that the singlet wave functions must be described 
by even Mathieu functions whereas the triplet ones must be described by odd Mathieu functions. 
The final form of the normalized singlet and triplet wave functions therefore is
\begin{align}
\psi_{kl}^{+}(R,z;q)&=\frac{\sqrt{2}}{L}\exp{\left(\frac{2\pi i k R}{L}\right)} C_l\left(z;q\right) 
\label{eq:kl_singlet}\\
\psi_{kl}^{-}(R,z;q)&=\frac{\sqrt{2}}{L}\exp{\left(\frac{2\pi i k R}{L}\right)} S_l\left(z;q\right)
\label{eq:kl_triplet}
\end{align}
in which the normalization of the Mathieu functions is chosen such that
\begin{equation}
\int\limits_{0}^{\pi}dz \left| C_l(z;q)\right|^2 = \int\limits_{0}^{\pi}dz \left| S_l(z;q)\right|^2 =\frac{\pi}{2} .
\label{eq:mathieu_norm}
\end{equation}
The Mathieu functions further have the periodicity property  $M_l(z+\pi)= (-1)^l M_l (z)$, i.e.
they are periodic in $\pi$ for even values of $l$ and anti-periodic for odd values of $l$. 
Furthermore the center-of-mass
wave function in Eq.(\ref{eq:ansatz}) changes with a prefactor $(-1)^k$ when $x_i \rightarrow x_i + L$, for $i=1,2$.
Therefore for the wave functions
in Eqs.(\ref{eq:kl_singlet}) and (\ref{eq:kl_triplet}) to satisfy periodic boundary conditions the labels $k$ and $l$ must be
both even or both odd. Note that $k$ runs over all integers while $l$ only runs over
the non-negative integers.
From Eq.(\ref{eq:mathieu_characteristic}) we see
that the energy eigenvalues are given by
\begin{equation}
E_{kl}^{\pm}(q)=\left(\frac{\pi}{L}\right)^2\left[k^2+a_l^{\pm}(q)+2q\right] .
\label{eq:energykl}
\end{equation}
Since in the subsequent discussion of the
density response function we focus on the singlet excitations in particular, we write
the singlet wave functions in a slightly different form for the purpose of a better interpretation. Multiplying the spatial
wave function of Eq.(\ref{eq:kl_singlet}) with its singlet spin function, the full space-spin function can be written
as
\begin{equation}
\Psi_{kl} (x_1 \sigma_1, x_2 \sigma_2) = \Phi_k (x_1 \sigma_1, x_2 \sigma_2)  \, \sqrt{2} \, C_l \left [ \frac{\pi}{L} \left( x_1-x_2 \right ) \right] 
\end{equation}
where we defined the Slater determinant
\[
\Phi_k (x_1 \sigma_1, x_2 \sigma_2) = \frac{1}{\sqrt{2}}  \left|  
\begin{array}{cc}\phi_{k/2} (x_1) \delta_{\sigma_1 \uparrow}  &  \phi_{k/2} (x_1) \delta_{\sigma_1 \downarrow} \\
 \phi_{k/2} (x_2) \delta_{\sigma_2 \uparrow}  &  \phi_{k/2} (x_2) \delta_{\sigma_2 \downarrow}  \end{array}
  \right|  
\]
and we further defined the spatial normalized orbital by $\phi_k (x) = e^{2 \pi i k x /L} /\sqrt{L}$ which corresponds to a periodic single-particle
wave function of a free particle on the quantum ring.
Let us consider the excitation from the ground-state to another singlet state with $l=0$, which requires that the excited state
is characterized by an even $k$ value. In that case $\phi_{k/2}$ in the Slater determinant above is a proper periodic wave function as $k/2$ is an integer. 
According to Eq.(\ref{eq:energykl})
the excitation energy is
\begin{equation}
\Delta E_{k0}^+ = E_{k0}^+ - E_{00}^+ = \left(\frac{\pi k}{L}\right)^2,
\label{eq:CM_excitations}
\end{equation}
which is independent of the interaction strength $q$ as the excited state has the same
relative wave function as the ground state. For the case that $q=0$ we have $C_0 (z;q=0)=1/\sqrt{2}$ and 
the ground and excited state both become pure Slater determinants. The excitation then represents a promotion of two electrons
from a doubly occupied $k=0$ state to a doubly occupied state with a one-particle quantum number $k/2$, which is commonly called a 
double excitation. When $q$ is non-zero this language is not accurate anymore as also the relative wave function
becomes relevant. If the interaction strength becomes very large the energy required to excite to a state
with non-zero $l$ becomes very large too and the excitations with energy $\Delta E_{k0}^+$ give the
dominant contribution to the density response function, as we will see later. 

\subsection{The strong interaction expansion of the exact solution}

As the interaction strength $q$ increases, the electronic repulsion becomes more important and the electrons tend to stay in opposite positions on the ring. 
This physically intuitive picture can be analyzed in more detail using the Mathieu equation.
According to Eqs. (\ref{eq:kl_singlet}) and (\ref{eq:kl_triplet}), the square of the spatial wave function is given by
\[
| \psi_{kl}^{\pm}(R,z;q) |^2 = \frac{2}{L^2} M_l^2 (z;q)
\]
where $M_l (z;q)$ is either a Mathieu cosine $C_l$ or a Mathieu sine $S_l$ depending on whether the  wave function is a singlet or a triplet one.
We therefore see that the probability to find a given electron at $x_2$ given an electron at $x_1$ only depends on the
relative coordinate $x_1-x_2$, as one would expect on the basis of the symmetry of the system. This probability distribution is given by the square of the Mathieu function $M_l$. 

\begin{figure}[ht]
  \centering
   \includegraphics[width=0.43\textwidth]{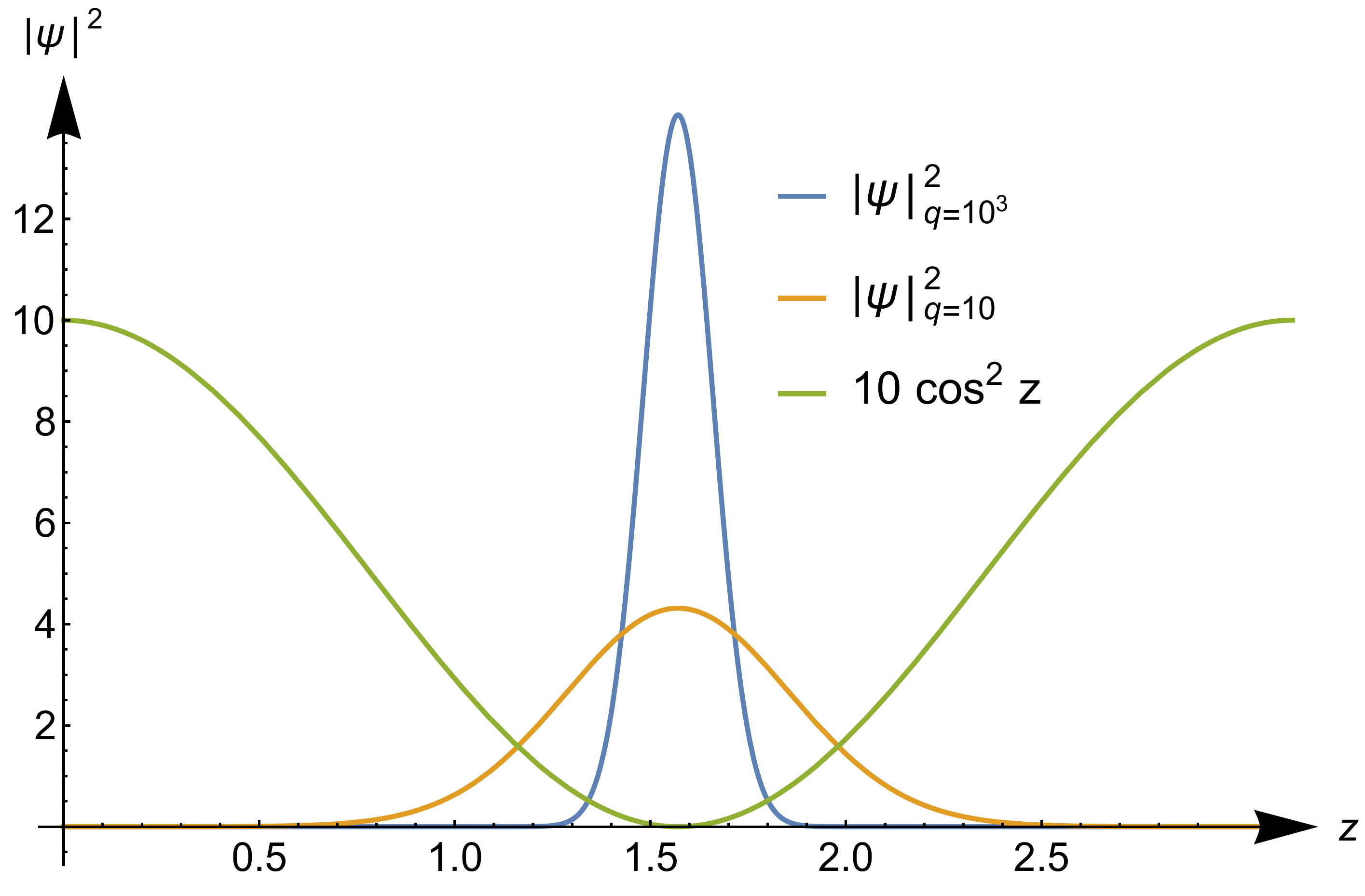}
  \caption{The squared ground-state wave function $|\psi_{00}|^2$ for two values of the interaction strength $q$, and the interaction $\cos ^2 z$ using a suitable scaling for showing it in the same plot. For large $q$, the wave function localizes around $z=\pi/2$ where $\cos^2z$ is almost parabolic and $|\psi_{00}|^2$  then  tends to a sharp Gaussian. 
  \label{WavePotential} }
\end{figure}

Let us analyze the properties of this function in the large interaction limit which according to
Eq.(\ref{eq:Mathieu}) satisfies a single-particle Schr\"odinger type of equation in a potential of the form 
$V(z)=2 q \cos (2z)$. For large values of $q$ we can see that the relative wave function described by $M_l$ becomes localized in the
minimum of the potential at $z=\pi/2$, which corresponds to a relative distance of the particles of $L/2$. We can expand the potential around this
minimum to obtain
\[
2q \cos (2z) = -2q + 4q \, (z - \frac{\pi}{2})^2 + \ldots
\] 
This potential describes (apart from a shift of the minimum) a harmonic oscillator with frequency $\Omega = 2 \sqrt{q}$.
The eigenfunctions of the harmonic oscillator are well known to consist of Gaussians of width proportional to $1/\sqrt{\Omega} =1 /(\sqrt{2} \, q^{1/4})$.
In the limit of large $q$ the harmonic frequency increases and the wave functions become localized around $z=\pi/2$.
This behaviour is illustrated in Fig.(\ref{WavePotential}).
The eigenenergies $\epsilon_l$ of the harmonic oscillator are well-known and given by  $\epsilon_l = -q + \Omega (l +1/2)=a/2$.
This also immediately provides an asymptotic formula for the characteristic value of the Mathieu equation for large values of $q$:
\[
a_l^{\pm} (q) = -2q + 2 \sqrt{q} \, (2l+1) + \ldots  
\]
and consequently also an asymptotic expansion for the eigenenergies of the quantum ring from Eq.(\ref{eq:energykl}).

A more rigorous connection to the harmonic oscillator wave functions can be made on the basis of the
substitution 
$u(z) = \sqrt{2} \, q^{1/4} \cos z$ which transforms the Mathieu equation (\ref{eq:Mathieu}) to the new form
 \begin{equation}
 \left[ -\frac{1}{2}(1- \frac{u^2}{2 \sqrt{q}}) \frac{d^2}{du^2} + \frac{u}{4 \sqrt{q}} \frac{d}{du} + \frac{u^2}{2} \right] \mathcal{M}(u)= \epsilon \,
 \mathcal{M}(u)
\nonumber
 \end{equation}
where we defined $\epsilon=(a+2q)/(4\sqrt{q})$ and $\mathcal{M}(u(z)) = M(z;q)$. In the large $q$ limit this
equation attains the form of the Schr\"odinger equation for the harmonic oscillator.
 Its eigenfunctions are well known and, apart from a normalization,  are
given by the parabolic cylinder functions
  $\mathcal{D}_m (u)$ defined by
 \begin{align}
 \mathcal{D}_m (u) &=\frac{(-1)^m}{2^{m/2}} e^{u^2/2} \frac{d^m}{du^m} e^{-u^2} \nonumber \\
 &= \frac{1}{2^{m/2}} e^{-u^2/2} H_m (u) 
 \label{eq:Dfunc}
 \end{align}
where $H_m (u)$ are the Hermite polynomials.
On the basis of this analysis we may suspect that it is possible to find an asymptotic large-$q$ expansion of the Mathieu functions in terms of harmonic oscillator functions $\mathcal{D}_m$ of argument $u$. 
Sips \cite{Sips1949,Sips1959,Blanch1960,Frenkel2001} already derived such an expansion on the basis of the transformed Mathieu equation. 
For reference in the next section we briefly outline its main features for the case of the Mathieu cosine which is relevant for the discussion of singlet states. 
The general form of the Sips expansion is given by
 \begin{equation}
 C_l (z;q) = \sum_{n=-\infty}^{\infty} c_{2n,l} (q) \, \mathcal{D}_{l+2n} \left [u(z) \right] 
 \label{eq:sips_cosine}
 \end{equation}
in which we defined $\mathcal{D}_{m<0}=0$.
The specific form of the coefficients $c_{2n,l} (q)$ is given in the work of Sips \cite{Sips1949,Sips1959,Blanch1960} who outlined a systematic procedure to obtain them.  In general they can be obtained from a recursion relation \cite{Frenkel2001} and we refer to Appendix~\ref{App:Sips} for a more detailed discussion.

In \Eq{eq:sips_cosine} we see that for odd values of $l$ the Mathieu cosine is expanded in functions $\mathcal{D}_m$ with only odd values of $m$ while for even $l$ it is expanded in functions $\mathcal{D}_m$ with only even values of $m$.  This follows directly from the derivation by Sips~\cite{Sips1949} but we see with hindsight that this condition is necessary to make the Mathieu cosine satisfy $C_l (z+\pi;q) = (-1)^l C_l (z;q)$. Namely, if we replace $z$ by $z+\pi$ then the variable $u$ changes to $-u$ 
yielding this desired property for a series of the form (\ref{eq:sips_cosine}) since $\mathcal{D}_m (-u) = (-1)^m \mathcal{D}_m (u)$. The Sips expansion, \Eq{eq:sips_cosine}, will be used in the next section to determine the large interaction expansion of the density response function.

\begin{figure}[ht]
  \centering
   \includegraphics[width=0.48\textwidth]{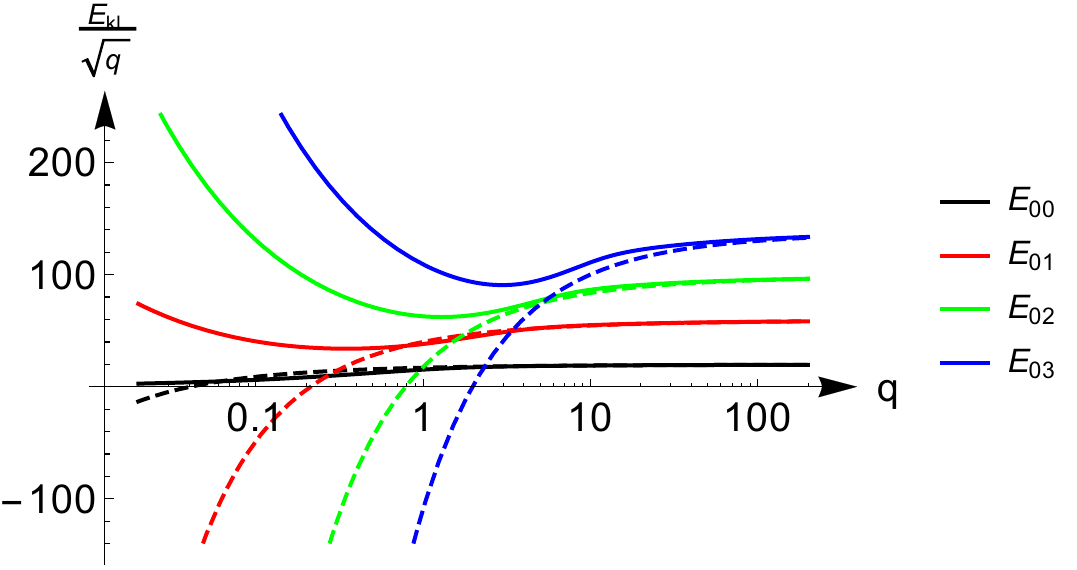}
  \caption{The ground-state $(l=0)$ and excited-state $(l=1,2,3)$ energies divided by $\sqrt{q}$ for $k=0$ as function of the interaction strength $q$ in units of $L^{-2}$. We plot the exact energies (solid lines) versus the approximate energies (dashed lines) from the expansion in Eq.(\ref{eq:asymp_kl}). 
  \label{ExcitedStates} }
\end{figure}

We conclude the section with a remark on the eigenenergies of the quantum ring. We can obtain an asymptotic expansion for them as the 
work of Sips also derives the large $q$ behavior of the
Mathieu characteristic values in terms of an asymptotic series expansion in power of $q^{1/2}$ (see Appendix~\ref{App:Sips}). Taking the
first few leading orders we obtain the following expression for the eigenenergies of the quantum ring
\begin{align}
 E_{kl}^\pm &= \! \left ( \frac{\pi}{L} \right )^2 \left [ k^2 + 2 \sqrt{q} \, (2l+1) - \frac{1}{4} (2l^2 +2 l+1) 
 \right . \nonumber \\
 &+ \left . \frac{(2l+1)}{128 \sqrt{q}}\left( (2l+1)^2+3 \right)
 \right ] + \ordo{q^{-1}}.
 \label{eq:asymp_kl}
\end{align}
The asymptotic expansion 
is the same for the singlet and triplet energies as their difference becomes exponentially small in the large $q$ limit (see Appendix~\ref{App:Sips}). To illustrate the $q$-dependence of the eigenenergies we present in Fig. \ref{ExcitedStates} some of the lowest eigenvalues and their asymptotic expansion from Eq.(\ref{eq:asymp_kl}) as a function of $q$. We see that the asymptotic expansion converges more slowly for higher values of $l$, and for these $l$ we need high values of $q$ in order to have a reliable estimate.

\section{Density response of strongly interacting electrons} \label{secDensityResponse}
After having discussed the two-particle wave function and energy spectrum of the system, let us now move to its response properties. Particularly relevant to TDDFT is the induced density change $\delta n(\rb,t)$ when a small time-dependent external potential $\delta v(\rb,t)$ is applied. 
They are related by the retarded density response function $\chi(\rb t,\rb' t')$ as follows
\begin{equation}
\delta n (\rb,t) = \int d\rb' \! \int dt' \, \chi (\rb t , \rb' t' ) \delta v(\rb',t'),
\label{eq:denschange}
\end{equation}
where $\chi$ is defined via 
\begin{align}
\chi(\rb t,\rb' t') & =\frac{\delta n(\rb, t)}{\delta v(\rb', t')} = \nonumber \\
& -i \theta (t-t') \langle \Psi_0 | [ \hat{n}_H (\rb,t), \hat{n}_H (\rb',t') ] | \Psi_0 \rangle 
\label{eq:dens_rep}
\end{align}
where $\hat{n}_H$ is the density operator in the Heisenberg picture and $\Psi_0$ is the ground state of the system. Since the unperturbed system is time-independent, the density response function is a function
of the relative time $\tau=t-t'$ only and we can Fourier transform it with respect to  $\tau$:
\[
\chi (\rb,\rb',\omega) = \int d\tau\, \chi (\rb,\rb',\tau) \, e^{i \omega \tau}.
\]
Before addressing in greater detail the properties of $\chi$ of the quantum ring in the large interaction limit, let us first make some considerations about the static density response function in a more general context. 

\subsection{Static density response in the strong interaction limit}

Let us consider an interacting many-electron system in its ground state. The Hamiltonian consists of a kinetic energy operator, an external potential $v(\mathbf{r})$ and a two-body interaction. 
If we consider a small variation $\delta v(\mathbf{r})$ in the static external potential, the ground-state density will vary by an amount $\delta n(\mathbf{r})$
which can be expressed as
\begin{equation}
\delta n(\mathbf{r}) = \int d\mathbf{r}' \, \chi (\mathbf{r}, \mathbf{r}') \, \delta v ( \mathbf{r}' )  
\label{eq:dn_dv}
\end{equation}
where $\chi(\rb,\rb') = \chi(\rb,\rb',\omega=0)$ is the static density response function.
Let us now consider a shifted potential $v' (\mathbf{r}) = v (\mathbf{r} + \mathbf{R} )$. 
The ground-state density for this new potential is given by $n' (\mathbf{r}) = n (\mathbf{r} + \mathbf{R} )$. For small translations we can write that $\delta n(\mathbf{r}) = n' (\mathbf{r}) - n(\mathbf{r}) = \mathbf{R} \cdot \nabla n (\mathbf{r})$ and similarly $\delta v(\mathbf{r}) =v' (\mathbf{r}) - v(\mathbf{r}) = \mathbf{R} \cdot \nabla v (\mathbf{r})$. Since this is valid for all small vectors $\mathbf{R}$
we find from Eq.(\ref{eq:dn_dv}) that
\begin{equation}
\nabla n (\mathbf{r}) =  \int d\mathbf{r}' \, \chi (\mathbf{r}, \mathbf{r}') \, \nabla' v ( \mathbf{r}' )  .
\label{eq:dndv_grad}
\end{equation}
This equation relates the gradient of the external potential to the gradient of the ground-state density, and amounts to the static limit of an equation derived for the dynamic density response function by Vignale~\cite{Vignale1995}. 

Let us now consider a system in which we scale the two-body interaction with a parameter $\lambda$ and let us choose the external potential $v_\lambda (\mathbf{r})$ in such a way that the density $n(\mathbf{r})$ is the same for all values of $\lambda$. According to the Hohenberg-Kohn theorem \cite{Hohenberg1964} such a potential is unique when it exists. For such a system the density response function will depend on $\lambda$ as well and Eq.(\ref{eq:dndv_grad}) becomes
\begin{equation}
\nabla n (\mathbf{r}) =  \int d\mathbf{r}' \, \chi_{\lambda} (\mathbf{r}, \mathbf{r}') \, \nabla' v_\lambda ( \mathbf{r}' )  .
\label{eq:dndv_grad2}
\end{equation}
Let us now consider the limit of very large values of $\lambda$. One can show, for a general inhomogeneous system, that asymptotically $v_{\lambda} (\mathbf{r}) = \lambda u (\mathbf{r}) +  \ldots$ where $u (\mathbf{r})$ is the so-called strictly-correlated electron potential \cite{Seidl1999,Seidl2007,Gori-Giorgi2009b}. The result is intuitively clear as the linearly growing repulsive two-body interaction must be compensated by a linearly growing attactive one-body potential in order to keep the density profile constant.
This has consequences for the behavior of $\chi_\lambda$. We consider two cases. Let us first assume that for $\lambda \rightarrow \infty$ the response function
$\chi_\lambda$ attains a finite value $\alpha (\mathbf{r},\mathbf{r}')$. Because the left-hand side of Eq.(\ref{eq:dndv_grad2}) is independent of
$\lambda$ this implies that
\begin{equation}
\mathbf{0} =  \int d\mathbf{r}' \, \alpha (\mathbf{r}, \mathbf{r}') \, \nabla' u ( \mathbf{r}' )  
\label{eq:alfa_func}
\end{equation}
which means that the three vector components of $\nabla u$ must be eigenfunctions of $\alpha$ with zero eigenvalue.
Since the density response function reaches a finite limit the system 
does not become rigid even when the interaction becomes infinitely large.
This is a possible situation in systems in which the Hamiltonian by a coordinate transformation 
can be separated in two parts in which one of the parts is weakly dependent on the interaction strength.
That such inhomogeneous systems exist is demonstrated for the harmonic model system described in Appendix~\ref{App:HarmonicExample} and for which we demonstrate that Eq.(\ref{eq:alfa_func}) is indeed valid.

A probably more common situation is that such a separation is either not possible or that both
parts of such a Hamiltonian are still strongly $\lambda$-dependent. In this case one would expect that the energy required to excite the system grows with $\lambda$, and as such the density response function would vanish for large $\lambda$.
If this is the case it is to be expected
that $\chi_\lambda$ in Eq.(\ref{eq:dndv_grad2}) asymptotically behaves as
\begin{equation}
\chi_{\lambda} (\mathbf{r}, \mathbf{r}') = \frac{1}{\lambda} \, \beta(\mathbf{r},\mathbf{r'}) + \ldots
\label{eq:beta}
\end{equation}
where $\beta$ is a $\lambda$-independent function and the terms that follow decay faster than $1/\lambda$.
This means that for a given perturbation $\delta v(\mathbf{r})$ the density response $\delta n(\mathbf{r})$
decays as $1/\lambda$ and therefore the strong interaction makes the system more rigid and suppresses density variations.
In such a case Eq.(\ref{eq:dndv_grad2}) reduces to
\begin{equation}
\nabla n (\mathbf{r}) =  \int d\mathbf{r}' \, \beta (\mathbf{r}, \mathbf{r}') \, \nabla' u ( \mathbf{r}' ),
\end{equation}
which is an exact equation for the leading order in $\lambda$.

When we finally consider systems in which the ground-state density and external potential
are spatially constant, the reasoning that we carried out does not apply anymore since the gradients in Eq.(\ref{eq:dndv_grad2}) are identically zero. However, such systems are 
homogeneous which implies that the center-of-mass can be separated off and we can therefore expect
the response function to attain a finite value in the large interaction limit.
This is exactly the case of our quantum ring model. Indeed we saw in Eq.(\ref{eq:CM_excitations})
that the quantum ring admits excitation energies that are independent of the interaction strength and these
correspond to excitations that only change the center-of-mass wave function and do not affect the relative probability distribution of the particles. As we will see in more detail below, such excitations give a contribution to the density response function that survives in the large interaction limit, while the remaining excitations give a contribution which behaves as in Eq.(\ref{eq:beta}).

It is interesting to connect this analysis to the $f$-sum rule for the dynamic density response function. In a system where the density is kept independent of $\lambda$ with $v_\lambda (\mathbf{r})$ the $f$-sum rule attains the form \cite{Stefanucci2013}
\begin{equation}
 \frac{1}{\pi} \int d\omega \, \omega\, \chi_\lambda (\mathbf{r},\mathbf{r}',\omega) = \nabla [ n (\mathbf{r}) \nabla \delta (\mathbf{r} - \mathbf{r}') ].
 \label{eq:f_sumrule}
\end{equation}
We therefore see that the frequency integration removes the $\lambda$-dependence. This is not in contradiction with Eq.(\ref{eq:beta}). Although the density response function itself can become very small for large $\lambda$ the integrand in Eq.(\ref{eq:f_sumrule}) can remain finite as it is weighted by the frequency $\omega$. As a consequence the integral gets contributions proportional to the excitation energies which grow with increasing interaction strength.

\subsection{Exact density response of the quantum ring}
After having discussed the general static case, we now turn our attention to the exact dynamical density response function of the quantum ring.
Inserting a complete set of eigenstates of the Hamiltonian $\hat{H}$ from \Eq{eq:Hamiltonian} into the one-dimensional analogue of Eq.(\ref{eq:dens_rep}) 
we find the Lehmann representation~\cite{Stefanucci2013} of the retarded response function
\begin{align}
\chi (x,x',\omega) \! =
\! \! \! \! \! \sum_{k,l,p=\pm} \! 
&\left[\frac{\langle\Psi_{00}^{+}|\hat{n}(x)|\Psi_{kl}^p\rangle\langle\Psi_{kl}^p|\hat{n}(x')|\Psi_{00}^+\rangle}{\omega-\Delta E_{kl}^p(q)+i\eta}\right.\nonumber\\
-&\left.\frac{\langle\Psi_{00}^+|\hat{n}(x')|\Psi_{kl}^p\rangle\langle\Psi_{kl}^p|\hat{n}(x)|\Psi_{00}^+\rangle}{\omega+\Delta E_{kl}^p(q)+i\eta}\right]\label{DefResponse}
\end{align}
where we defined the excitation energies as $\Delta E_{kl}^p(q) = E_{kl}^p(q)-E_{00}^{+}(q)$.
The expression contains an infinitesimal parameter $\eta>0$ that arises from the Fourier transform of the Heaviside function and the limit $\eta \rightarrow 0$ is implied
after the evaluation of all terms. Furthermore $\hat{n}(x)$ is the density operator in the Schr\"odinger picture and $p=\pm$ labels the singlet or triplet eigenstates.
The label $k$ runs over all positive and negative integers while $l$ runs over non-negative integers, with the condition that both are even or both are odd.
The expression in \Eq{DefResponse} is simplified by the fact that the triplet terms vanish because the triplet spin-function is orthogonal to the singlet spin function of the ground state, which yields $\langle\Psi_{00}^+|\hat{n}(x)|\Psi_{kl}^-\rangle=0$. 
The remaining non-zero terms can be evaluated as
\begin{align}
\langle\Psi_{00}^+|\hat{n}(x_1 )|\Psi_{kl}^+\rangle &
= 2 \int_0^L dx_2 \,  \psi_{00}^{+*}  (x_1,x_2) \psi_{kl}^{+} (x_1,x_2) \nonumber \\
& = \frac{2}{L} e^{2\pi i k x_1/L}  D_{kl}(q)  \label{eq:transSym}
\end{align}
where $\psi_{kl}^+$ denotes the spatial part of the singlet wave function of
Eq.(\ref{eq:kl_singlet}) expressed in the original coordinates and the excitation amplitudes $D_{kl}(q)$ read
\begin{align}
D_{kl}(q)&=\frac{2}{\pi}\int\limits_{0}^{\pi}dz\ C_{0}(z;q)C_{l}(z;q)e^{-ikz}.
\label{dintegral1}
\end{align}
The amplitude $D_{kl}(q)$ has a number of properties directly related to properties of the Mathieu functions.
Since $C_{l}(z;q)$ is real, $D_{kl}^{*} (q) = D_{(-k)l} (q) $, and
as a consequence of the orthogonality of the Mathieu functions, $D_{0l} (q) = \delta_{l0}$. 
Moreover, the fact that $C_l (z+\pi;q) =(-1)^l  C_l (z;q)$ and that these functions are even in $z$ leads to $D_{kl} (q)=(-1)^{k+l} D_{kl}^{*} (q)$.
Making use of the symmetry properties of $D_{kl}(q)$ described, combined with $\Delta E_{kl}^+ (q) = \Delta E_{(-k)l}^+ (q)$, yields 
the following expansion of the response function
\begin{align}
\chi(x,x', \omega) =& \frac{1}{L} \sum_{k=-\infty}^{\infty}  \chi(k,\omega) \, e^{2\pi i k \left ( x - x' \right ) /L} 
\label{eq:chi_exp}
\end{align}
where 
\begin{align}
& \chi(k,\omega) = \nonumber \\ 
& \frac{4}{L}  \sum_{l}  \left( \frac{|D_{kl} (q)| ^2}{\omega - \Delta E_{kl}^+ (q)+ i\eta} - 
\frac{|D_{kl} (q)| ^2}{\omega + \Delta E_{kl}^+ (q) + i\eta} 
\right) \nonumber \\
&= \frac{8}{L} \sum_{l} \frac{ \Delta E_{kl}^+ (q) \ |D_{kl}(q)|^2}{(\omega+i\eta)^2- (\Delta E_{kl}^{+}(q))^2}
\label{eq:chi_k}
\end{align}
in which the sum runs over even values of $l$ for $k$ even and over odd values of $l$ for $k$ odd.
We see from Eq.(\ref{eq:chi_exp}) that $\chi (k,\omega)$ can be regarded as the discrete Fourier transform
of $\chi (x,x',\omega)$ with respect to the relative spatial coordinate $x-x'$ as was to be expected on the basis
of the symmetry of the system.
It will be now convenient to define
the spatially discrete and temporally continuous Fourier transform of a function $f(x,t)$ and its inverse as:
\begin{align}
f(k,\omega) &=  \int_0^L dx \, e^{ - \frac{2 \pi i k}{L} x } \int_{-\infty}^\infty dt \, e^{ i \omega t } f(x,t) \\
f(x,t) &= \frac{1}{L} \sum_{k=-\infty}^{\infty} e^{\frac{2 \pi i k}{L}x} \int_{-\infty}^\infty \frac{d\omega}{2 \pi } \, e^{-i \omega t} f (k,  \omega) .
\end{align}
By using this Fourier transformation in Eq.(\ref{eq:denschange})
we rewrite the density response as
\begin{equation}
\delta n (k,\omega) = \chi (k,\omega) \delta v (k,\omega),
\end{equation}
in which $k$ is an integer and $\omega$ a continuous variable.
We will make use of this relation below. 
We have now obtained an explicit form of the density response function that allows for an analytical analysis in the
strong interaction limit. 

However, before moving to that, we briefly give the form of the response function for
the non-interacting system, i.e. $q=0$, which in the density functional context will be the same as the Kohn-Sham
response function, since the system has the same density for all values of $q$. For the noninteracting case
the Mathieu characteristic value is $a_l^+ (0)=l^2$ and the Mathieu cosine 
functions are given by $C_0 (z;0) = 1/\sqrt{2}$ and $C_l (z;0) = \cos (lz)$ for $l \geq 1$. 
The excitation energies are given by \Eq{eq:energykl}, 
\begin{equation}
E_{kl}^{+}(0)=\left(\frac{\pi}{L}\right)^2\left[k^2+l^{2} \right] , \nonumber
\label{eq:energykl_nonint}
\end{equation}
while the eigenstates are given by Eq.(\ref{eq:kl_singlet}) as 
\begin{align}
& \psi^+_{kl} (x_1,x_2) = \nonumber \\
& \frac{1}{\sqrt{2}} [ \phi_{\frac{k+l}{2}} (x_1) \phi_{\frac{k-l}{2}} (x_2)  + \phi_{\frac{k+l}{2}} (x_2) \phi_{\frac{k-l}{2}} (x_1) ]
\end{align}
in which $k \pm l$ is always even. Note that $l \neq |k|$ yields a doubly excited state. The corresponding excitation amplitude can be calculated from Eq.(\ref{dintegral1}). 
Apart from the amplitude $D_{00} (0)=1$,
which does not contribute to the Lehmann sum since $\Delta E_{00}^+ =0$, for $(kl) \neq (00)$ we have that
\begin{equation}
D_{kl} (0) = \left\{ \begin{array}{cc}  1/\sqrt{2}   &   \mbox{if $l = |k|$}  \\ \frac{\sqrt{2}}{\pi}  \frac{i k}{k^2-l^2} [(-1)^{k+l}-1]  & \mbox{if $l \neq |k|$.} 
\end{array} \right. \label{DklNoninteracting}
\end{equation}
Since only terms where $k+l$ is even contribute, we see that the only non-zero excitation amplitudes are the ones with $l=|k|$. 
This implies the absence of double excitations in the density-response function, a well-known property 
of non-interacting systems~\cite{Maitra2004,Ullrich2012}.  Inserting \Eq{DklNoninteracting} into Eq.(\ref{eq:chi_k}) 
we find that the
non-interacting response function $\chi_s (k,\omega)$ is given by
\begin{equation}
\chi_s (k,\omega) = \frac{4}{L}  \frac{ \Delta E_{kk}^+ (0) }{(\omega+i\eta)^2- (\Delta E_{kk}^{+}(0))^2}
\label{chi0}
\end{equation}
with $\Delta E_{kk}^+ (0) = 2 (\pi k /L)^2$. We note that in $k$-space, $\chi_s$ has only a single pole for $\omega>0$, and no zeroes.

Having determined the non-interacting response function, it now remains to study the density response function in the complementary limit of very strong interactions.
For this purpose we need to study the excitation energies $\Delta E_{kl}^+ (q)$ and excitation amplitudes $D_{kl} (q)$ in the limit of large $q$. This is the topic of the next section.

\subsection{Strong interaction expansion of the dynamic density response function}

Let us focus on the excitation energies $\Delta E_{kl}^+ (q)$ and the excitation amplitudes $D_{kl} (q)$ for large $q$. Because for $\Delta E_{kl}^+ (q)$ explicit 
asymptotic expansions are known (see Appendix~\ref{App:Sips}), this leaves us with the determination of $D_{kl} (q)$ defined by Eq.(\ref{dintegral1}).
We start by inserting the Sips expansion of Eq.(\ref{eq:sips_cosine}) into \Eq{dintegral1}, which gives
\begin{equation}
D_{kl} (q) = \frac{2}{\pi} \sum_{n_1,n_2 = -\infty}^\infty c_{2n_1,0} (q) c_{2n_2,l} (q) \mathcal{J}_{kl}^{n_1 n_2} (q)
\label{eq:Dkl_exp}
\end{equation}
where we defined
\begin{equation}
 \mathcal{J}_{kl}^{n_1 n_2} (q) = \int_0^\pi dz \, e^{-ikz} \mathcal{D}_{2n_1} (u) \mathcal{D}_{2n_2+l} (u) 
\end{equation}
where $u(z) = \sqrt{2} \, q^{1/4} \cos z$. Since the coefficients $c_{2n,l} (q)$ are known (see Appendix~\ref{App:Sips} for explicit expressions) it remains to evaluate $\mathcal{J}_{kl}^{n_1 n_2} (q)$. 
Changing the integration variable to $u$ and defining $b=\sqrt{2} q^{1/4}$ gives the expression
\begin{equation}
 \mathcal{J}_{kl}^{n_1 n_2} (q) = \frac{1}{b} \int_{-b}^{b} du \, f_k \left ( \frac{u}{b} \right ) \, \mathcal{D}_{2n_1} (u) \mathcal{D}_{2n_2+l} (u) 
 \label{eq:J2}
\end{equation}
where we defined the function
\begin{equation}
f_k (x) = \frac{e^{-ik \arccos (x)}}{\sqrt{1-x^2} } = \sum_{r=0}^{\infty} a_r (k) \, x^r
\label{eq:fkx}
\end{equation}
and its Taylor coefficients $a_r (k)$. Inserting this Taylor series into Eq.(\ref{eq:J2}) then gives the expansion
\begin{equation}
 \mathcal{J}_{kl}^{n_1 n_2} (q) = \sum_{r=0}^\infty \frac{a_r (k)}{b^{r+1} }  \int_{-b}^{b} du \, u^r \, \mathcal{D}_{2n_1} (u) \mathcal{D}_{2n_2+l} (u)
 \nonumber
\end{equation}
where the interchange of integral and sum is allowed as we have an absolutely convergent series.
Due  to the Gaussian decay of the functions $\mathcal{D}_n (u)$,  in the limit $q \rightarrow \infty$, we make an error which, as a function of $q$, decays faster than any polynomial function if we replace $b$ in the limits of the integral
by infinity. We therefore obtain the asymptotic expansion
\begin{equation}
 \mathcal{J}_{kl}^{n_1 n_2} (q) = \sum_{r=0}^\infty \frac{a_r (k)  I_{n_1 n_2, r}^l
}{(\sqrt{2} q^{1/4})^{r+1} }
\label{eq:Jcoeff}
\end{equation}
where we introduced coefficients of the form
\begin{equation}
I_{n_1 n_2, r}^l =  \int_{-\infty}^\infty du \, u^r \, \mathcal{D}_{2n_1} (u) \mathcal{D}_{2n_2+l} (u). \label{eq:theI}
\end{equation}
This integral can be computed analytically, and the explicit expression is given in Appendix~\ref{App:expansion}. Also note that due to the parity properties of the integrand $I_{n_1 n_2, r}^l$ 
vanishes unless $r$ and $l$ are both even or both odd.
Therefore, depending on whether $l$ is even or odd, the summation index $r$ in Eq.(\ref{eq:Jcoeff}) can be taken to run only over even or only over odd values. 

\begin{figure*}[ht]
  \centering
   \includegraphics[width=0.30\textwidth]{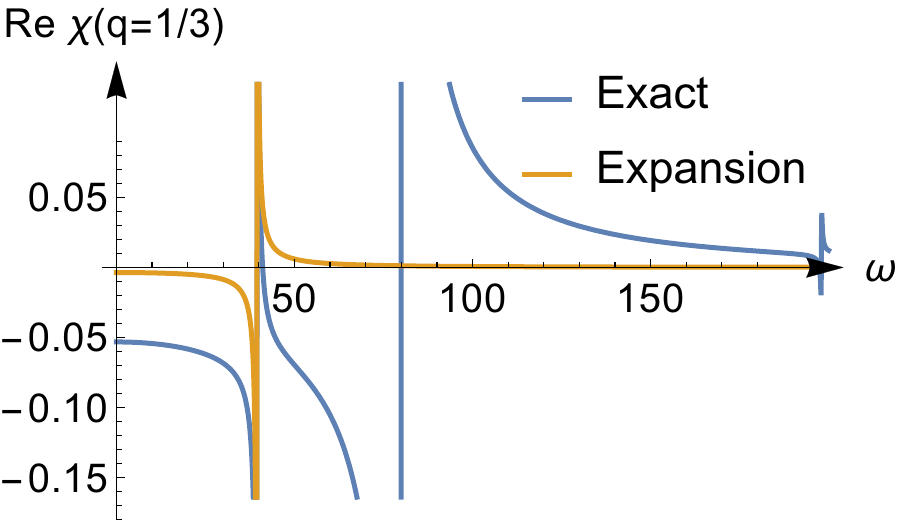}
   \includegraphics[width=0.30\textwidth]{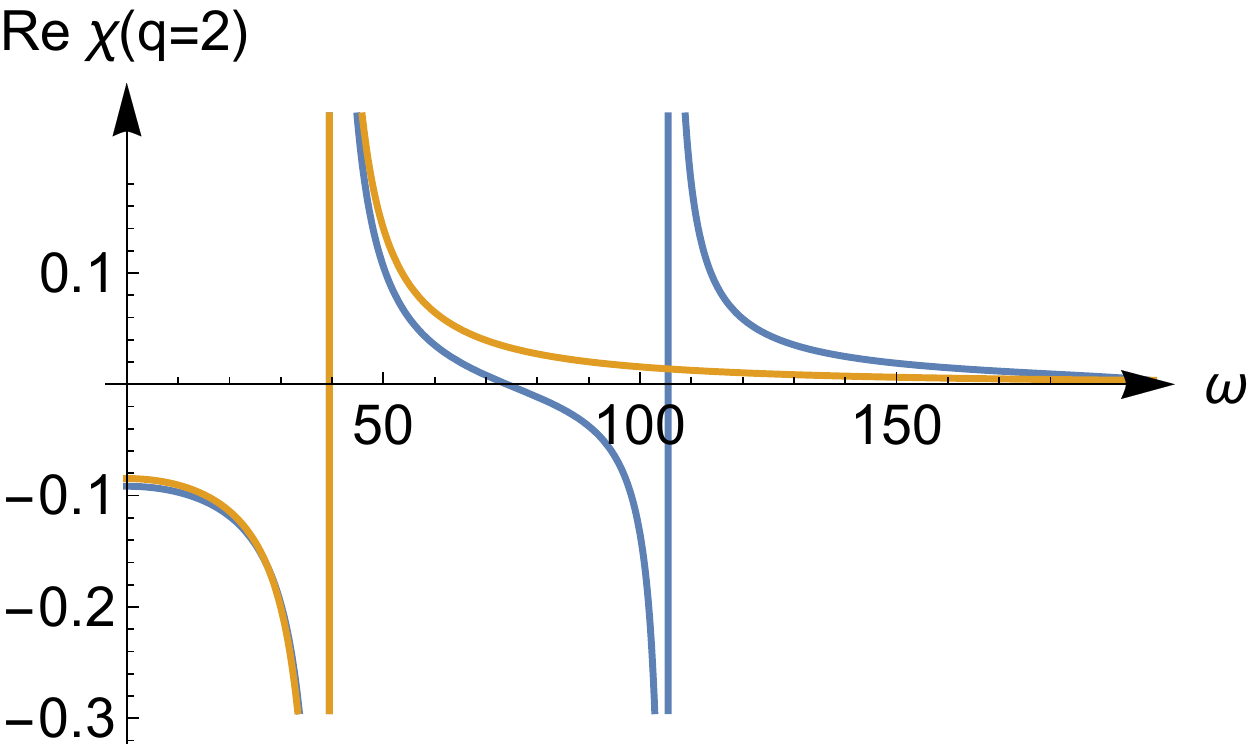}
   \includegraphics[width=0.30\textwidth]{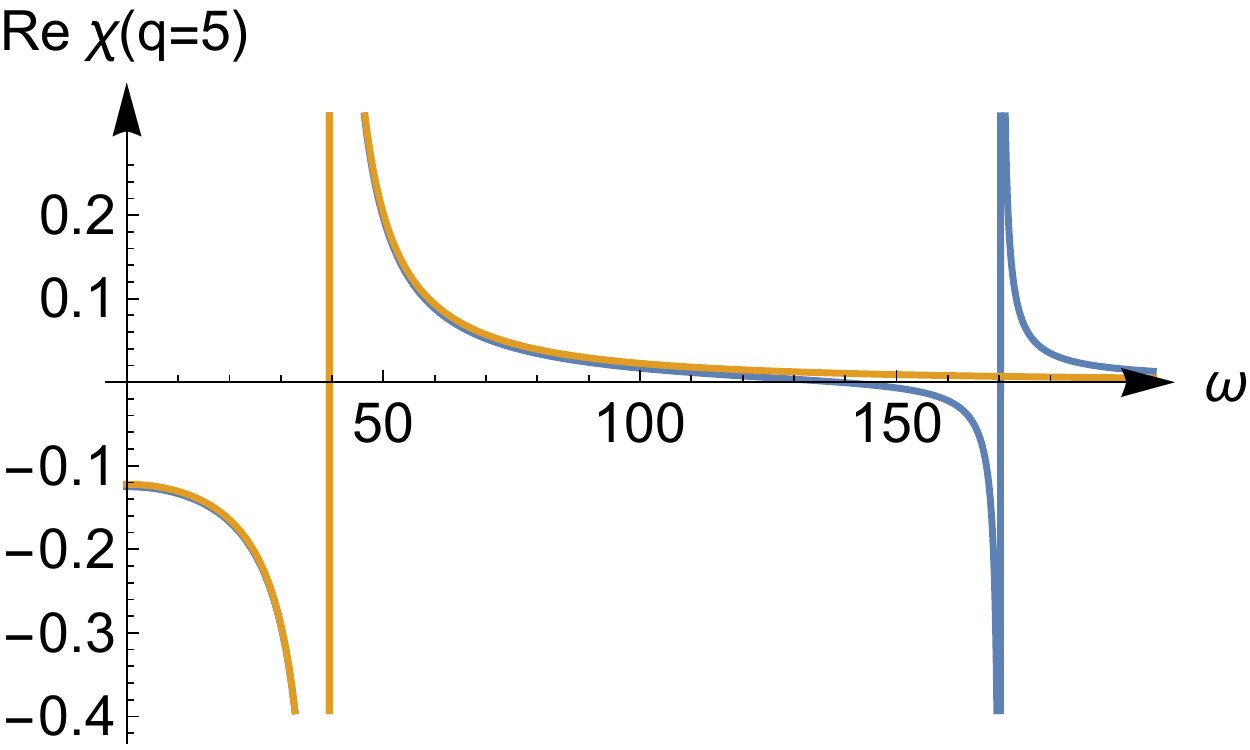}
   \includegraphics[width=0.30\textwidth]{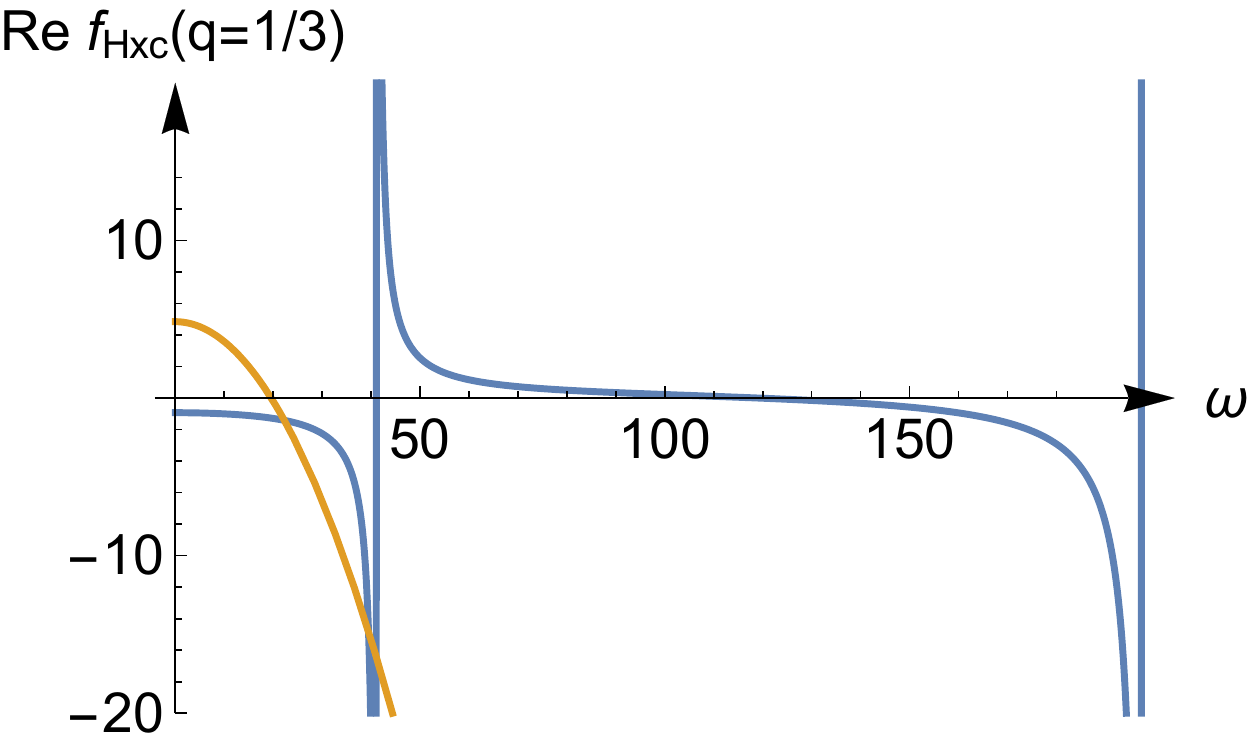}
   \includegraphics[width=0.30\textwidth]{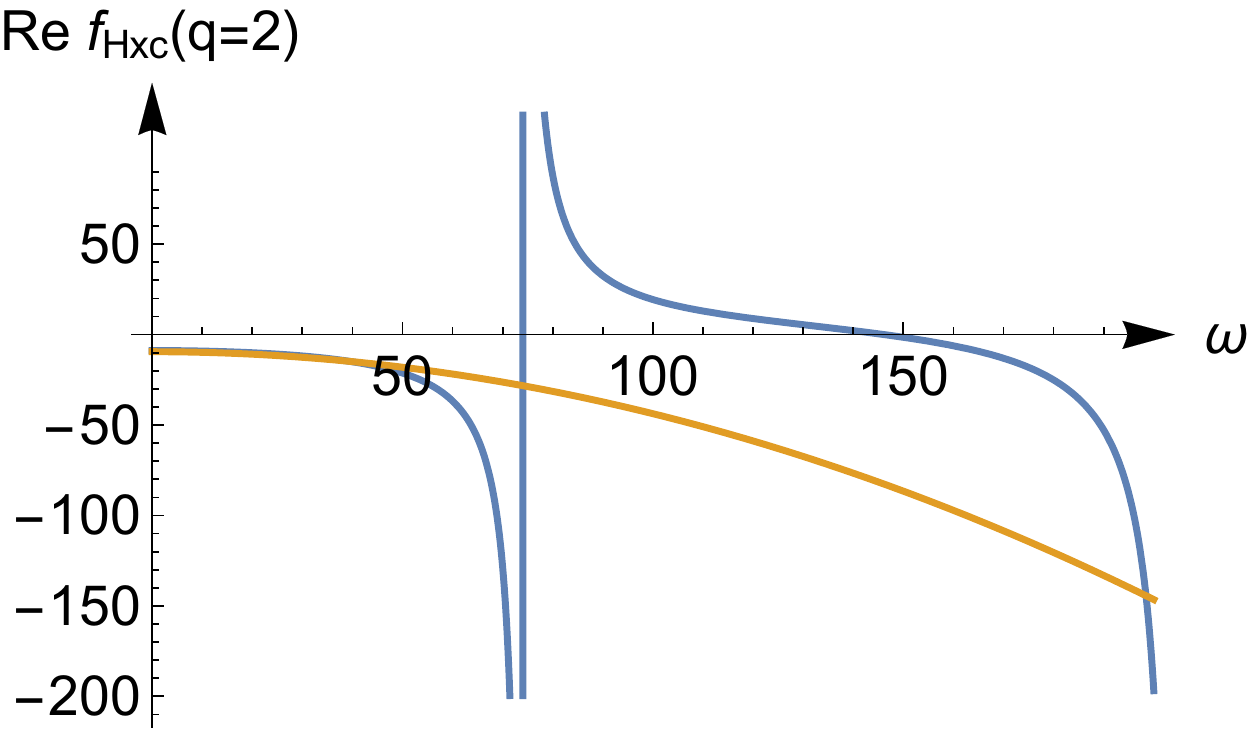}
   \includegraphics[width=0.30\textwidth]{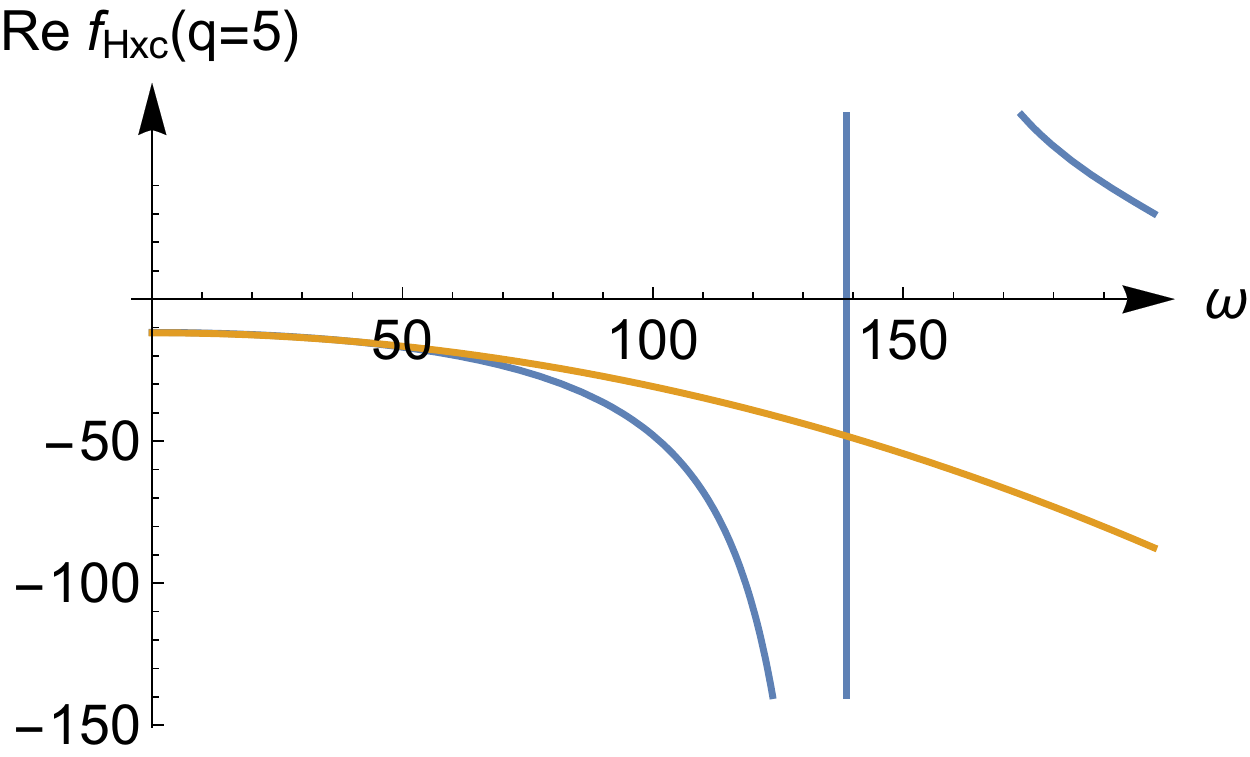}
  \caption{The real part of $\chi(k=2,\omega)$ in units of $L$ (top) and $f_{\Hxc}(k=2,\omega)$ in units of $L^{-1}$ (bottom) for different values of the interaction strength $q$. The exact results are obtained by numerical integration, and the expansion of $\chi$ and $f_{\Hxc}$ is taken up to the same order as shown in 
  Eq. (\ref{eq:resp_qlim}) and Eq. (\ref{fhxc_fourier}).   \label{chiandfxcFigures} }
\end{figure*}

Expression (\ref{eq:Jcoeff}) together with Eq.(\ref{eq:Dkl_exp}) gives an explicit procedure to calculate the large $q$ expansion of the excitation amplitudes. The asymptotic expansions of $D_{kl}(q)$ and of $\left | D_{kl}(q) \right |^2$ are given in Appendix~\ref{App:expansion} in \Eq{eq:Dkl} and \Eq{eq:D2} respectively. Together with the asymptotic expansion for the excitation energies, \Eq{eq:excitationEnergies}, inserted into \Eq{eq:chi_k} we find that the asymptotic expansion of the density response function is given by
\begin{widetext}
\begin{equation}
\chi (k,\omega) = \left\{ \begin{array}{ll}  \frac{8}{L}\frac{( \pi k/L)^2}{(\omega+i\eta)^2-\left(\frac{\pi k}{L}\right)^4} 
\left[1 - \frac{k^2}{4 \sqrt{q}} + \frac{k^2 (k^2-2)}{32 q} \right] + \mathcal{O}(q^{-3/2}) &  \mbox{if $k$ is even} \\ 
\frac{L}{2 \pi^2} \left[  - \frac{k^2}{q} + \frac{k^2 (k^2-1)}{2 q^{3/2}}  -  \frac{k^2}{q^2} \left( \frac{368 k^4 -928 k^2 +947}{2304}  + \frac{1}{16} \left( \frac{L}{\pi}\right)^4  \omega^2 \right) \right] + \mathcal{O} (q^{-5/2}) & \mbox{if $k$ is odd.} \end{array} \right.
\label{eq:resp_qlim}
\end{equation}
\end{widetext}

From this expression, we can draw a number of interesting conclusions. 
We find that the response function behaves quite differently for even and odd Fourier coefficients in the strong interaction regime.
If we apply a potential with general coefficients $\delta v(k,\omega)$ to the system, the density change $\delta n(k,\omega)$ is strongly suppressed for odd $k$  as $\chi (k,\omega)$ becomes very small. 
In this limit it will therefore mainly have even Fourier coefficients which implies that
$\delta n (x,t) = \delta n (x +\frac{L}{2},t)$
i.e. the density change at antipodal points of the ring is the same. If we however apply a
potential with only odd Fourier coefficients the density change has the symmetry
$\delta n (x,t) = -\delta n (x +\frac{L}{2},t)$
and is therefore opposite in antipodal points of the ring.
From Eq.(\ref{eq:resp_qlim}) we find that the leading term in real space in this case is given by
\begin{equation}
\delta n (x,t ) = \frac{n_0}{q}  \left( \frac{L}{2 \pi}\right)^4  \partial_x^2 \delta v(x,t).
\label{eq:dn-odd}
\end{equation}

We can understand the dependence on $q$ as follows.
The generation of an anti-symmetric antipodal density requires excitation to states 
with an odd number of nodes in the relative wave function
which requires a large energy in the strong interaction limit and therefore the density response is
suppressed for large interaction strength.
From Eq.(\ref{eq:dn-odd}) we also see that the density increases instantaneously around the points were the potential has positive curvature.
The instantaneous nature of the response has a simple explanation.
If we perturb the system with a potential $\delta v(k,\omega)$ which is only non-zero
for frequencies $\omega$ well below the first excitation energy, the temporal variation of the
perturbation is much slower than a typical timescale of the free evolution of the system and the
density response can be regarded as instantaneous.
Since the excitation energies $\Delta E_{kl}^+$ for odd $k$ (which must have odd $l$ as well) increase proportionally to
$\sqrt{q}$ the density response function in this case is well approximated by a frequency
independent function for $\omega \ll \sqrt{q}$, which explains the instantaneous dependence of the 
density variation on the perturbation in Eq.(\ref{eq:dn-odd}). 

For even values of $k$ the density response function has a more interesting frequency dependence.
In the strong interaction limit
\begin{equation}
\lim_{q \rightarrow \infty} \chi (k,\omega) =
  \frac{8}{L}\frac{( \pi k/L)^2}{(\omega+i\eta)^2-\left(\frac{\pi k}{L}\right)^4}.
  \label{chi_large}
\end{equation}
The poles of this response function correspond to the center-of-mass excitations of
Eq.(\ref{eq:CM_excitations}). Being independent of $q$ they are not shifted towards infinity when we increase the interaction strength.
This is a peculiarity of the quantum ring system as the Hamiltonian is separable in a $\lambda$-dependent and a $\lambda$-independent part. 
This happens also for some other homogeneous systems such as the three-dimensional electron gas with periodic boundary conditions or for electrons restricted to the surface of a sphere \cite{Seidl2007a}.
The analysis based on Eq.(\ref{eq:dndv_grad2}) shows that such a separation is usually not possible
in inhomogeneous systems. When it is possible, both parts will generally still depend on $\lambda$ (see Appendix~\ref{App:HarmonicExample} for an example).

To illustrate the accuracy of the expansion in Eq.(\ref{eq:resp_qlim}) 
we display the exact response function and the expanded one in the top panels of Fig.\ref{chiandfxcFigures} for $k=2$ and for some values of the interaction strength.
For small interactions $(q=1/3)$ the exact response function has two poles; one is approximately at the same location as the Kohn-Sham response function $(\omega = 2 (\pi k /L)^2)$, while a new pole with a small weight appears at the center-of-mass excitation energy at $\omega = (\pi k/L)^2$. The expansion captures this pole, albeit with a very different weight. When we increase the interaction strength, the pole originally at the Kohn-Sham energy will shift to higher energies to a position proportional to $\sqrt{q}$, while the pole corresponding 
to the center-of-mass 
excitation stays fixed and increases in weight. Already at $q=5$, the asymptotic expansion yields good results for this $k$-value. 
We thus see that the expansion is accurate for frequencies that are small compared to $\sqrt{q}$. 
This result was to be expected since in the expansion of Eq.(\ref{eq:resp_qlim}) we treated the frequency $\omega$ as a constant that is small compared $\sqrt{q}$. 

Having obtained the exact response function we have obtained all information needed to study the xc-kernel of TDDFT, which will be the topic of the next section.

\section{TDDFT in the strong interaction limit} \label{secTDDFTStrong}

\subsection{The exchange-correlation kernel}

\label{sec:qr-kernel}

In TDDFT an effective non-interacting system, known as the Kohn-Sham system,
is constructed in such a way as to have exactly the same density as the interacting many-particle system of interest. The external
potential in this system, $v_s ([n];\br t)$, is a functional of
the density \cite{Ruggenthaler2013,Ruggenthaler2015} and is often written as follows
\begin{equation}
v_s (\br t) = v(\br t) +  \int d\br' \, w (\br,\br') n (\br',t) + v_{\textrm{xc}} (\br t).
\label{eq:KSpot}
\end{equation}
Here $v(\br t)$ is the external potential of the interacting system of interest and $w (\br,\br')$ the two-particle interaction of that system.
The second term in Eq.(\ref{eq:KSpot}) is the Hartree potential and the last one is the 
exchange-correlation (xc) potential. 
Taking the functional derivative of Eq.(\ref{eq:KSpot}) with respect to the density one obtains
\begin{equation}
\chi_s^{-1} (\br t, \br' t') = \chi^{-1} (\br t, \br' t') + f_{\Hxc} (\br t , \br' t') .
\label{eq:fhxc}
\end{equation}
Here $\chi_s^{-1}$ is the inverse of the Kohn-Sham density response function whereas
$\chi^{-1} $ is the inverse of the density response function of the interacting system and $f_\Hxc$, the
Hartree-xc kernel, is defined as  
\begin{equation}
f_\Hxc (\br t , \br' t')  = \frac{\delta v_\Hxc (\br t )}{\delta n (\br' t')},
\label{fhxc_def}
\end{equation}
where $v_\Hxc$ is the sum of the Hartree and the xc-potential. Equation (\ref{eq:fhxc}) is commonly used to calculate $\chi$ from the knowledge of $\chi_s$ at the price of approximating $f_\Hxc$.

For the discussion in the next section it is important
to  note that the functional derivative is not a uniquely defined function \cite{Hellgren2012,Ruggenthaler2013} due to the fact that for a system with a fixed number of particles the
density change must integrate to zero at any time, i.e.
\begin{equation}
0 = \int d \br \, \delta n (\br t).
\label{dens}
\end{equation}
Let us define a new function
\begin{equation}
\tilde{f}_\Hxc (\br t , \br' t')  = f_\Hxc (\br t , \br' t') + g (\br,t,t') + h (\br',t,t') 
\label{fhxc_tilde}
\end{equation}
with $g$ and $h$ arbitrary functions. The change in the
Hartree-xc potential produced by the kernel of Eq.(\ref{fhxc_tilde}) due to a density change $\delta n$ is given by
\begin{align}
\delta \tilde{v}_\Hxc (\br t) &= \int d\br' dt' \tilde{f}_\Hxc (\br t, \br' t' ) \delta n (\br' t') \nonumber \\
&= \int d\br' dt' f_\Hxc (\br t, \br' t' ) \delta n (\br' t') \nonumber \\
& +  \int d\br' dt'  [g (\br,t,t') + h (\br',t,t')  ] \delta n (\br' t')  \nonumber \\
&= \delta v_\Hxc (\br t ) + C(t)
\end{align}
where the integral over $g$ integrates to zero as a consequence of Eq.(\ref{dens})
and the integral over $h$ yields
a function $C(t)$ of time $t$ only, which is merely a gauge of the potential.
We therefore see that $\tilde{f}_\Hxc$ and $f_\Hxc$ are physically equivalent integral kernels.
The quantity that is defined unambiguously~\footnote{Note that Ref.~\cite{Lani2016} did not discuss the possible addition of arbitrary functions of one spatial variable to the kernel.} is the mixed spatial derivative
\begin{equation}
\nabla_\br \nabla_{\br'}  \tilde{f}_\Hxc (\br t , \br' t')  = \nabla_\br \nabla_{\br'}  f_\Hxc (\br t , \br' t'), 
\end{equation}
a property that will be used below.
 
Let us turn to the specific case of the quantum ring.
The density response function is diagonal in the momentum-energy representation and for the Fourier components the following relation holds:
\begin{equation}
f_\Hxc (k,\omega) = \frac{1}{\chi_s (k,\omega)} - \frac{1}{\chi (k,\omega)} .
\label{fhxc_qr}
\end{equation}
By Fourier transforming the kernel $f_\Hxc$ we impose a dependence on the relative coordinate in real space, which reduces the ambiguity of Eq.(\ref{fhxc_tilde})
to that of adding an arbitrary spatially constant function. In Eq.(\ref{fhxc_qr}) this freedom is reflected in the fact
that the kernel is well-defined for all $k$-values except for $k=0$, since in this case both the response functions vanish.
For the homogeneous quantum ring the Kohn-Sham response function coincides with the response
function of truly non-interacting  electrons of Eq.(\ref{chi0}). Using this equation, together
with the expansion of Eq.(\ref{eq:resp_qlim}), we obtain an explicit expression
for $f_\Hxc$ in the strong interaction limit:
\begin{widetext}
\begin{align}
\displaystyle
 f_{\Hxc}(k,\omega) = 
 \begin{cases}
  -\frac{3 \pi ^2 k^2}{8 L}
  -\frac{1}{\sqrt{\lambda}} \frac{L^2}{\sqrt{ V_0}} \frac{1}{16\pi} \left[ \omega ^2 - \left (\frac{\pi k}{L}\right )^4  \right]  +
  \frac{1}{\lambda} \frac{L}{ V_0}  \frac{k^2+2}{64}   \left[ \omega ^2 - \left (\frac{\pi k}{L}\right )^4  \right]  + \mathcal{O} (\lambda^{-3/2})  & \text{if $k$ is even} \\ \\
  \lambda \, \frac{ V_0 L }{2 k^2} + 
  \sqrt{\lambda} \, \frac{\sqrt{V_0} \pi  \left(k^2-1\right) }{2 k^2} 
  -\frac{\pi ^2 \left(368 k^4+224 k^2+371\right)}{1152 k^2 L} + \mathcal{O} \left ( \lambda^{-1/2} \right ) &
  \text{if $k$ is odd}
 \end{cases}
 \label{fhxc_fourier}
\end{align}
\end{widetext}
where we have reintroduced the variable $\lambda$ rather than $q$ as the $\lambda$ notation is commonly used in the density functional context, 
which will be central for the discussion in the next section. Since these quantities only differ by a numerical prefactor (see Eq.(\ref{eq:q_def})) we will refer
to the large interaction regime as the regime in which either of these two variables tends to infinity.

To illustrate the accuracy of the expansion in Eq.(\ref{fhxc_fourier}) we
display the exact $f_\Hxc$ kernel and the expanded one in the bottom panels of Fig.\ref{chiandfxcFigures} for $k=2$ and some values of the interaction strength.
 We see that the corresponding asymptotic expansion for $f_{\Hxc}$ is accurate up to the lowest excitation energy corresponding to a change in the relative wave function.
In this energy region $f_\Hxc$ has a pole at an energy corresponding to a zero in $\chi$, as a consequence of Eq.(\ref{fhxc_qr}). It is worth noticing that, since 
the non-interacting response function has no zeroes, there is no pole in $f_\Hxc$ originating from the first term in Eq.(\ref{fhxc_qr}).
This is peculiar to our quantum ring system for which the non-interacting 
response function has only a single pole for $\omega>0$. Instead, for a general system,
the Kohn-Sham response function will have multiple poles and zeroes which implies that, to order $\lambda^0$, $f_\Hxc$ is frequency dependent causing the adiabatic approximation to fail
in this order.  
In our system $f_{\Hxc}$ for even $k$ tends to a frequency-independent function for all $\omega$ when the interaction strength approaches infinity. Its static value, 
given by  $f_{\Hxc} = -\frac{3L}{8} \left ( \frac{\pi k}{L} \right )^2=\chi_s^{-1} \left [k,\omega=\left (\frac{\pi k}{L} \right )^2 \right]$, is the value needed to shift the Kohn-Sham pole to the $q \to \infty$ pole.  

For the odd $k-$values (not shown here), all poles in the response function shift to infinity as $q \to \infty$.
The asymptotic expansion for $f_\Hxc$ captures this, and the kernel becomes frequency-independent in this limit.
In fact, for odd values of $k$ in Eq.(\ref{fhxc_fourier}), all the leading terms up to order $\lambda^{0}$
are frequency independent. However, frequency-dependent terms will appear to order $\lambda^{-1/2}$ (not presented here) as is
also the case for even $k$.

For the discussion in the next section it is useful to recast the kernel in real space.
The expressions in Eq.(\ref{fhxc_fourier}) are sufficient to calculate this quantity 
to order $\lambda^0$ in real space using the Fourier transform of Eq.(\ref{F-trans})  
\begin{equation}
f_\Hxc (r,\omega) = \frac{1}{L} \sum_{k=-\infty}^\infty f_{\Hxc} (k,\omega) e^{\frac{2\pi  i k r}{L}}
\label{F-trans}
\end{equation}
and $r=x-x'$ is the relative distance between the points $x$ and $x'$
which are the one-dimensional counterparts of the spatial points in Eq.(\ref{fhxc_def}).
Since $x$ and $x'$ are both in the interval from $0$ to $L$ we have that $r \in [-L,L]$.
We find that
\begin{equation}
f_\Hxc (r,\omega) = \lambda f_1 (r ) +  \sqrt{\lambda} f_2 (r) + f_3 (r) + \mathcal{O} (\lambda^{-1/2})
\label{fhxc_exp}
\end{equation}
The leading term is given explicitly by
\begin{equation}
f_1 (r) = \frac{V_0 \pi^2}{2L} \left[ - \left|r \right| + \left|r + \frac{L}{2} \right| +  \left|r - \frac{L}{2} \right| -\frac{3L}{4} \right]
\label{f_1}
\end{equation} 
in which we choose the arbitrary constant function (see the discussion below Eq.(\ref{fhxc_qr})) such that the Fourier coefficient of $f_1$ becomes zero for $k=0$.
For $f_2$ we find (up to a constant) that
\begin{align}
f_2 (r) =  \frac{ \sqrt{V_0} \pi}{4}  \left[  \delta (r) - \delta \left ( r + \frac{L}{2} \right ) - \delta \left(r - \frac{L}{2} \right)\right] \nonumber \\
- \frac{\sqrt{V_0} \pi^3}{2 L^2}   \left[ - |r| + \left|r + \frac{L}{2} \right| +  \left| r - \frac{L}{2} \right| -\frac{3L}{4} \right] .
\label{f_2}
\end{align}
Since in the next section we will focus mostly on $f_1$ and $f_2$, we do not report here the real space representation of $f_3$.
In the next section we will show how $f_1$ and $f_2$ can be calculated in an alternative manner using the SCE theory in
the adiabatic approximation.

\subsection{Expanding the xc-kernel in the theory of strictly correlated electrons}
To date, no good and reliable approximations
for electrons in the strong correlation regime have been developed within TDDFT. A ground-state theory of so-called strictly correlated electrons (SCE) \cite{Seidl1999,Seidl2007}  has been constructed and applied within the adiabatic approximation to calculate the xc-kernel to the leading order in the interaction strength.
In this section we will first benchmark this approximation against the exact solution for the quantum ring model, and then derive and compare the next order.

Let us begin with a brief overview of the ingredients of SCE theory that we will use. The Hartree-xc energy for a system with interaction strength $\lambda$ can be written as \cite{Langreth1975}:
\begin{equation}
E_\Hxc^\lambda = \int^\lambda_0 d\lambda' \, W_{\lambda'} [n]
\label{exclambda}
\end{equation}
where we defined:
\begin{equation}
W_\lambda [n] = \langle \Psi_\lambda [n] | \hat{W}  |   \Psi_\lambda [n]  \rangle.
\end{equation}
In this expression  $\hat{W}$  is the two-particle interaction and $\Psi_\lambda [n] $ is the ground-state wave function of a system with a local external potential, interaction $\lambda \hat{W}$ and ground-state density $n$.
In the strong interaction limit  $W_\lambda [n]$ can be expanded as\cite{Gori-Giorgi2009b},\footnote{Strictly speaking the absence of a term of order $\lambda^{-1}$ in this expansion has only been shown for a Coulomb system but we find it to be true in our model as well}:
\begin{equation}
W_\lambda [n] = V_\textrm{SCE}[n] + \frac{V_\textrm{ZPE}[n]}{\sqrt{\lambda}} + \ordo{\lambda^{-3/2}}
\label{wmu}
\end{equation}
where the first term is the SCE energy which has the explicit form:
\begin{equation}
V_\textrm{SCE}[n]  = \frac{1}{2} \sum_{i=2}^N \int d\br \, n(\br)  w(| \br - \mathbf{f}_i ([n];\br)  |)
\end{equation}
where $w$ is the two-particle interaction which we assume to depend only on the distance between the particles. The functions $\mathbf{f}_i ([n];\br)$ are the so-called co-motion functions, which specify the position of $N-1$ electrons given the position of one electron at $\br$.
The second term in Eq.(\ref{wmu}) contains the so-called zero-point energy (ZPE) which describes the vibrations of the electrons around their equilibrium positions and is given explicitly as:
\begin{equation}
V_\textrm{ZPE}[n] = \frac{1}{2} \int d\br \frac{n(\br)}{N} \sum_{n=1}^{D(N-1)} \frac{\omega_n (\br)}{2}
\label{eq:ZPE}
 \end{equation}
where $D$ is the spatial dimensionality of the system and $\omega_n $ are the harmonic frequencies. 
Inserting Eq.(\ref{wmu}) into Eq.(\ref{exclambda}) we find the large $\lambda$ expansion of the Hartree-xc energy to be:
\begin{equation}
E_\Hxc^\lambda = \lambda V_\textrm{SCE}[n]  + 2 \sqrt{\lambda} V_\textrm{ZPE}[n] + \ordo{\lambda^0}.
\label{elambda}
\end{equation}
The ground-state theory can be used in the adiabatic approximation~\cite{Lani2016} to find an approximate exchange-correlation kernel from
\begin{equation}
f^{\textrm{A}}_\Hxc (\br t ,\br'  t') = \frac{\delta^2 E_\Hxc^\lambda}{\delta n (\br) \delta n (\br')} \, \delta (t-t')
\end{equation}
where we added the superscript $\textrm{A}$ to indicate that we make the adiabatic approximation. This approximation
 yields a frequency independent Hartree-xc kernel when transformed to frequency space, which we denote as $f^\textrm{A}_\Hxc (\br,\br')$.
 The second order variation of Eq.(\ref{elambda}) with respect to the density gives an expansion in orders of $\sqrt{\lambda}$ for the Hartree-xc kernel:
\begin{equation}
f^\textrm{A}_\Hxc (\br,\br') = \lambda \, f^{\textrm{ASCE}} (\br, \br') + \sqrt{\lambda} \, f^{\textrm{AZPE}} (\br, \br') + \ordo{\lambda^0}
\end{equation}
where we defined the adiabatic SCE and ZPE kernels as
\begin{align}
f^{\textrm{ASCE}} (\br, \br') &= \frac{\delta^2 V_\textrm{SCE}}{\delta n (\br) \delta n (\br')} 
\label{ASCE} \\
f^{\textrm{AZPE}} (\br, \br') &= 2 \frac{\delta^2 V_\textrm{ZPE}}{\delta n (\br) \delta n (\br')}.
\label{AZPE}
\end{align}
Let us now turn again to the case of the quantum ring. In this case there are two electrons and just one simple co-motion function
$\frm: [0,L] \rightarrow [0,L]$ given by
\begin{align}
\frm(x) = 
\begin{cases}
 x + \frac{L}{2}  &  \mbox{if $x \in [0,\frac{L}{2} [ $}  \\ 
 x - \frac{L}{2}  &  \mbox{if $x \in [\frac{L}{2} ,L ] $.} 
\end{cases}
\end{align}

If one electron is at $x$ this function simply puts the other electron at the antipodal point of the quantum ring. From this co-motion function it is straightforward to calculate the SCE energy. Since $|x-\frm (x)|=L/2$ we have from our interaction $w(x) = V_0 \cos^2 (\pi x/L)$ in Hamiltonian (\ref{eq:Hamiltonian}) that $w(|x-\frm (x)|)=0$ and therefore $V_\textrm{SCE}=0$ for our density.
Physically this means that the electrons are simply localized at the bottom of a potential well with zero energy. 

To next order oscillations around these equilibrium positions start to appear. These zero-point oscillations give an energy contribution which can be calculated using Eq.(\ref{eq:ZPE}). For a one-dimensional two-electron system there is only one non-zero harmonic frequency 
given by \cite{Malet2014}
\begin{equation}
\omega_1 (x) = 
\sqrt{ w'' ( |x-\frm (x)|) \left( \frac{n(x)}{n \left [\frm (x) \right]} +  \frac{n \left[\frm (x)\right]}{n(x)} \right) },
\label{eq:harmonic}
 \end{equation}
 where $w''(x) = \partial^2_x w(x)$.
If we calculate this frequency for our quantum ring we find $\omega_1 (x)= 2\pi \sqrt{V_0}/L$ and $V_\textrm{ZPE}=\pi \sqrt{ V_0} / (2L)$.
We can verify that this is in accordance with the exact strong interaction expansion of the Hartree-xc energy. 
Since the Kohn-Sham kinetic energy as well as the external potential is zero for the quantum ring system, $E_\Hxc^\lambda$ simply coincides with the total energy, which is known in the strong interaction limit from the large-$q$ expansion of the lowest Mathieu characteristic value (see Appendix~\ref{App:Sips}). This gives
\begin{equation}
E_\Hxc^\lambda =  \frac{\pi \sqrt{V_0}}{L}  \sqrt{\lambda}   - \frac{\pi^2}{4 L^2} - \frac{\pi^3}{16 L^3 \sqrt{V_0}} \frac{1}{\sqrt{\lambda}}  + \ordo{\lambda^{-1}}
\end{equation}
where the leading coefficient indeed exactly gives $2 V_\textrm{ZPE}$. The expansion of $W_\lambda$ can be calculated
from Eq.(\ref{exclambda}) to give
\begin{equation}
W_\lambda = \frac{dE_\Hxc^\lambda}{d\lambda}=\frac{\pi \sqrt{V_0}}{2L} \frac{1}{\sqrt{\lambda}} + \frac{\pi^3}{32 L^3 \sqrt{V_0}} \frac{1}{\lambda^{3/2}} + \ordo{\lambda^{-2}}.
\end{equation}
This result agrees with a direct calculation of $W_\lambda$ using
the Sips expansion of the Mathieu functions and indeed has the structure of the expansion in Eq.(\ref{wmu}) in which we also included the term to order $\lambda^{-3/2}$.

Let us turn to the calculation of the kernels of Eqs.(\ref{ASCE}) and (\ref{AZPE}).
\begin{figure}[ht]
  \centering
   \includegraphics[width=0.43\textwidth]{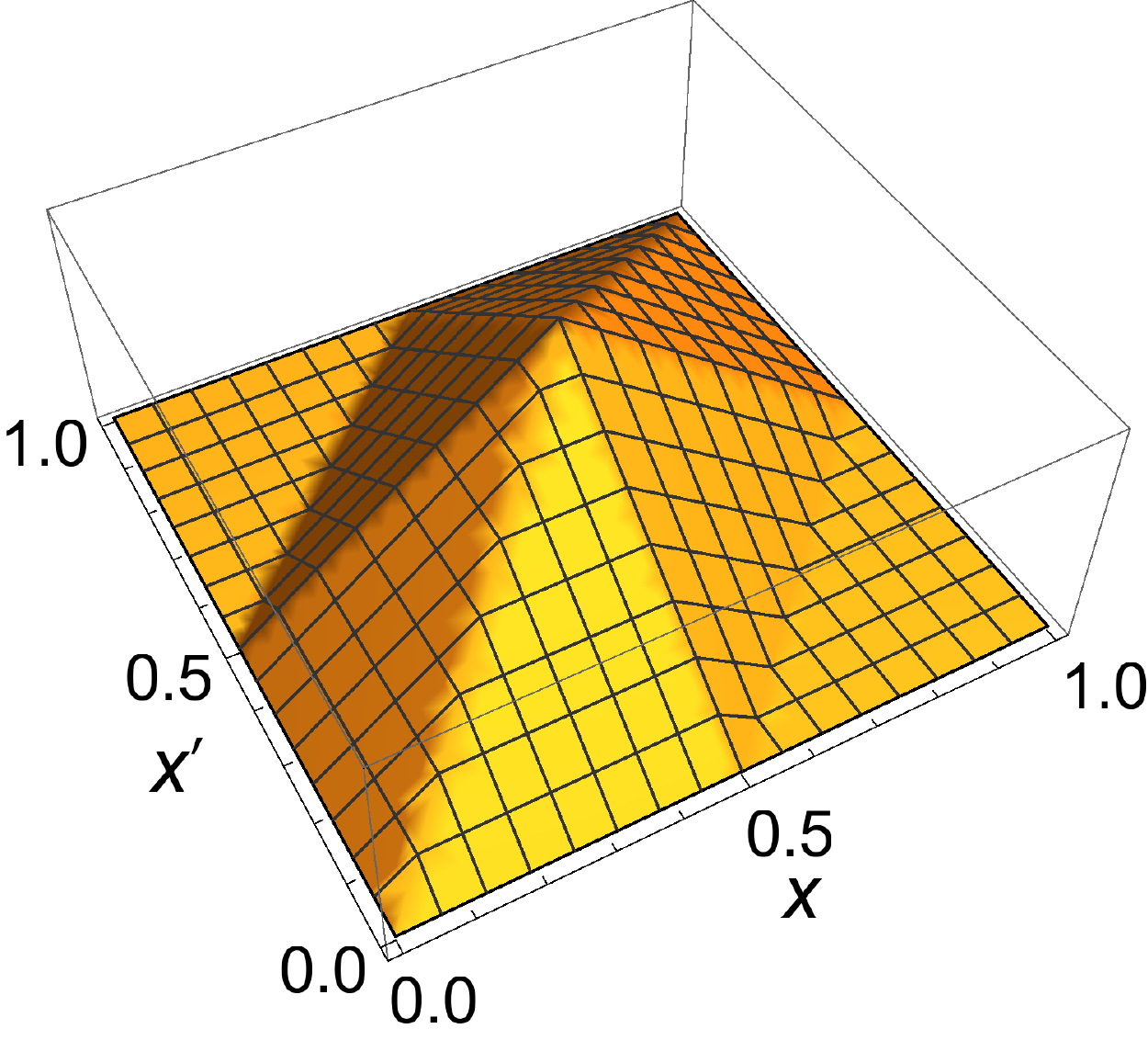}
   \includegraphics[width=0.43\textwidth]{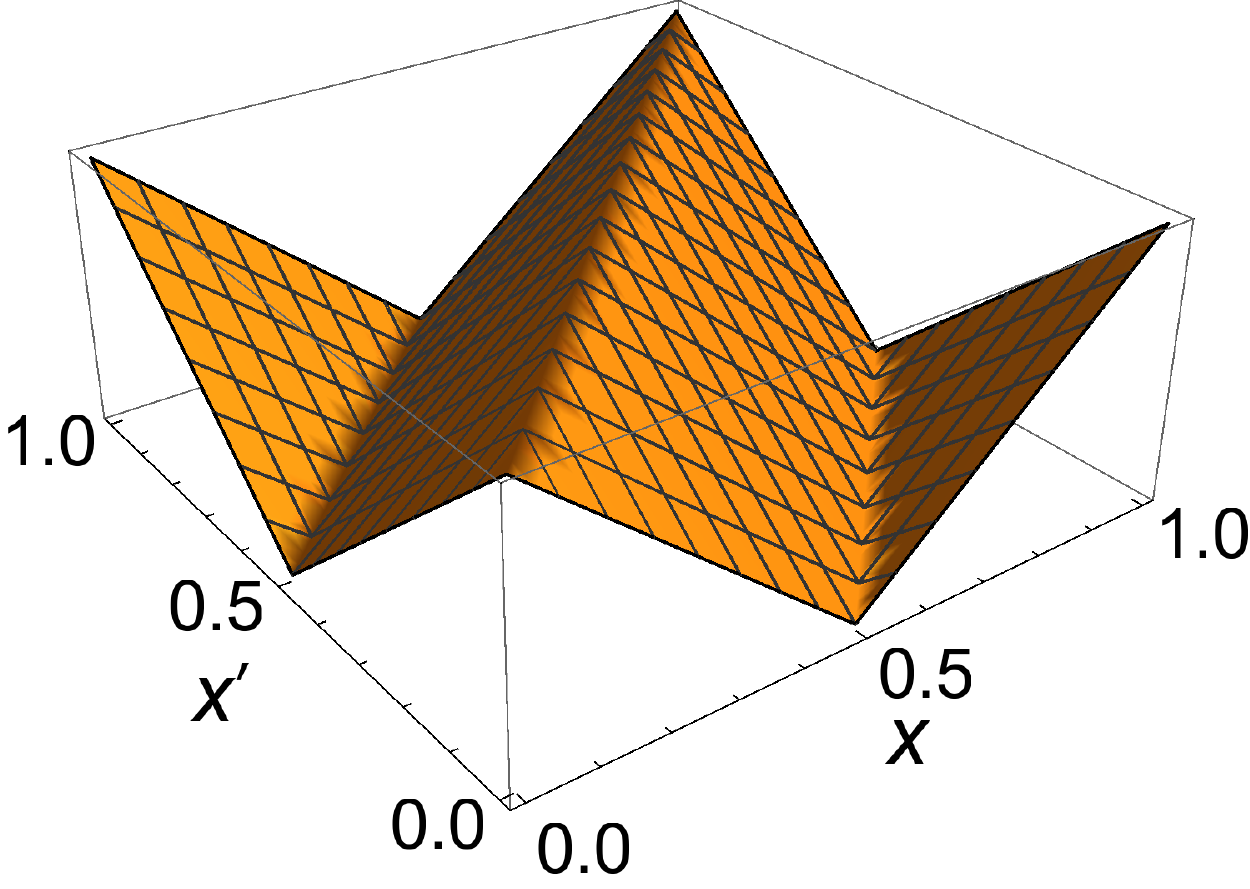}
  \caption{Top: The Hxc kernel $f^\textrm{ASCE}(x,x')$ of Eq.(\ref{eq:f1_kernel}). Bottom: The physically equivalent Hxc kernel $f_1 (x-x')$ of Eq.(\ref{f_1}). The $x$ and $x'$-axes are in units of $L$ and $f_{\Hxc}$ is given in arbitrary units.
  \label{fig:fxcTogether} }
\end{figure}
The ASCE kernel is obtained from the expression derived in Ref.~\cite{Lani2016} 
and reads
\begin{align}
&f^\textrm{ASCE}(x,x') = \nonumber \\
&- \int_0^x dy \frac{w^{\prime \prime}(|y-\frm (y)|)}{n \left [\frm (y) \right ]}\left[ \theta ( y-x' ) - \theta (\frm (y)-x' ) \right] \nonumber \\
&= f_1 (x-x') - f_1 (x) - f_1 (x') + f_1 (0)
\label{eq:f1_kernel}
\end{align}
where $f_1$ is the function given in Eq.(\ref{f_1}).
Because of the freedom in Eq.(\ref{fhxc_tilde}) $f^\textrm{ASCE}$ is physically equivalent to the kernel $f_1 (x-x')$ and agrees with the leading term in
the expansion of Eq.(\ref{fhxc_exp}) in the strong interaction limit. 
Both kernels are shown in Fig.\ref{fig:fxcTogether}.
Since $f_1(x-x')$ has a simpler shape, we will restrict ourselves to this function. The kernel describes how a density variation induces a change in $v_{\Hxc}$. To leading order in $\lambda$ we have
\begin{align}
\delta v_\Hxc (x,t)= \lambda 
\int dx'  f_1 (x - x') \delta n (x',t). \nonumber
\end{align}
By taking the second derivative of this equation, we obtain 
\begin{align}
&\partial^2_x  \delta v_\Hxc (x,t) = \lambda \frac{V_0 \pi^2}{L} \nonumber \\
&\times \left [-\delta n (x,t) +\delta n (x+L/2,t) + \delta n (x-L/2,t) \right ],
\label{eq:nonlocal}
\end{align}
where we stress that $\delta n (x,t)$ is periodic with $L$. Since $f_1(x-x')$ is linear everywhere except at the kinks at $x-x' = 0,\pm L/2$, the second derivatives yield delta functions at these points. Also note that the sign of the density change yields the curvature of the induced potential. Eq.(\ref{eq:nonlocal} ) has some interesting consequences.
If we make a localized density variation $\delta n (x,t)$ in a very small interval of the ring, there will not only be a change $\delta v_\Hxc$ in the same interval, but at the same time a similar change in the potential with opposite sign in
an antipodal interval. This shows very clearly that the Hxc-potential depends non-locally, but instantaneously, on the density. 

Let us analyze the next orders in the strong interaction expansion. 
The calculation of $f^{\textrm{AZPE}}$ allows for a comparison with the next leading term in Eq.(\ref{fhxc_exp}) which is proportional to $\sqrt{\lambda}$. 
The kernel $f^{\textrm{AZPE}}$ was obtained by taking the second functional derivative of the one-dimensional counterpart of Eq.(\ref{eq:ZPE}), using Eq.(\ref{eq:harmonic}) and the functional derivative of the
co-motion functions (derived already in Ref.~\cite{Lani2016}). 
We also find agreement between this expression and that of the next leading term in Eq.(\ref{f_2}), i.e. $f^{\textrm{AZPE}} (x,x')=f_2 (x-x')$ modulo the addition of arbitrary functions of $x$ and $x'$ separately (see again Eq.(\ref{fhxc_tilde})).  
We observe that the first two leading terms of the expansion of the Hartree-xc kernel from the adiabatic SCE theory agree with the exact results for the quantum ring. An interpretation of this fact will be presented below. 

\begin{figure}[ht]
  \centering
   \includegraphics[width=0.48\textwidth]{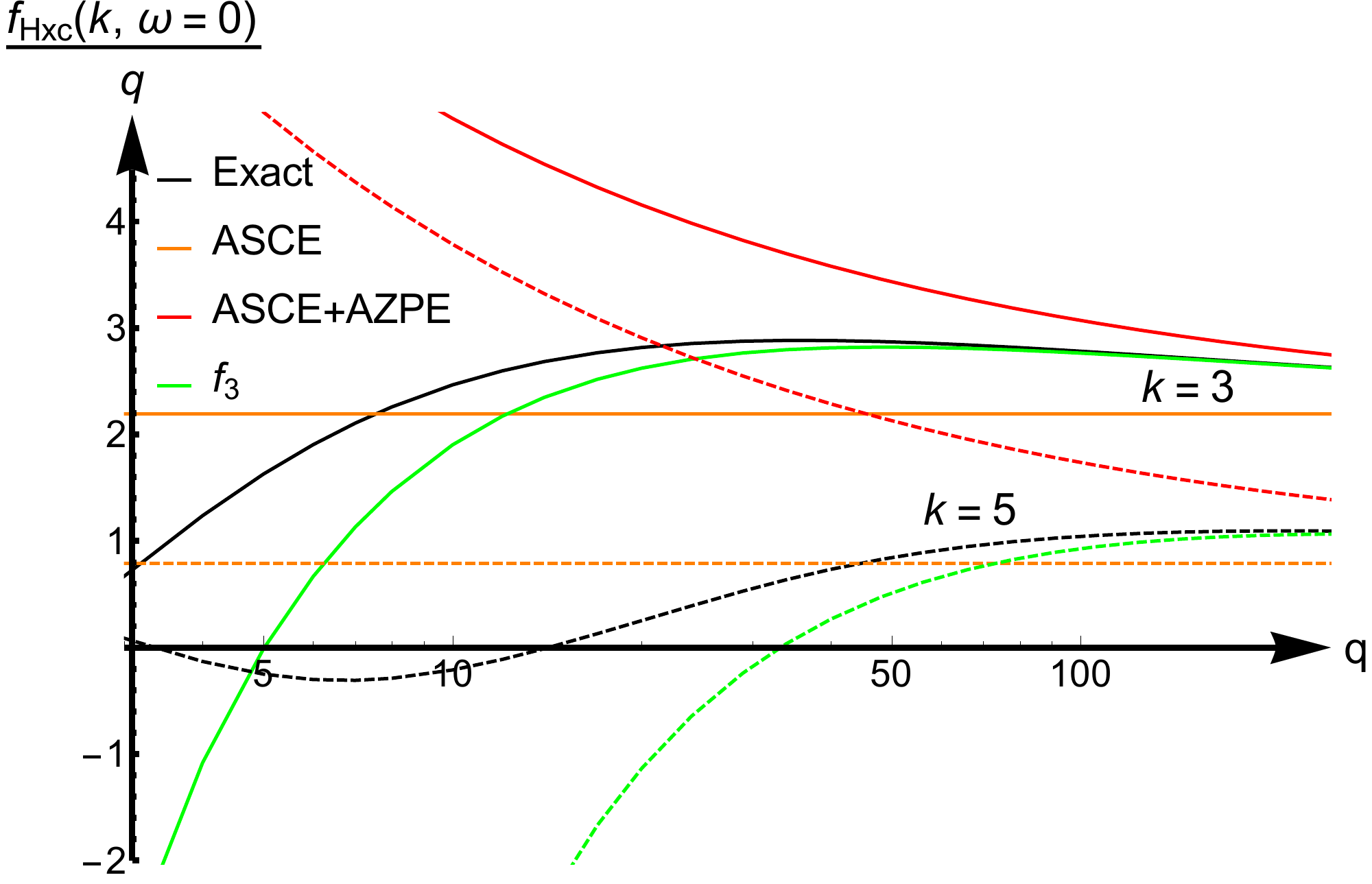}
   \caption{The static $f_{\Hxc}(k,\omega=0) / q$ as function of $q$, for $k=3$ (solid lines) and $k=5$ (dashed lines) in units of $L^{-1}$. We show the exact kernel obtained by numerical integration, and compare ASCE, ASCE+AZPE and ASCE+AZPE+$f_3$ (denoted by $f_3$ in the figure) 
   coming from \Eq{fhxc_exp}. Note that we plot the kernel as function of $q$ instead of $\lambda$ (see \Eq{eq:q_def}) in order to be consistent with the previous figures. 
  \label{fxcqFigure} }
\end{figure}

We have thus seen that, in this model, the ASCE and AZPE terms agree with the terms $f_1$ and $f_2$ respectively of the exact asymptotic expansion.
To better elucidate their role in the strong interaction limit, we show in Fig. \ref{fxcqFigure} the first three terms  contributing to $f_{\Hxc}$, all scaled by $\lambda$, and compare them with the expression for the exact kernel, in Fourier space for $k=3$ and $k=5$. 
As was pointed out earlier the accuracy of the expansion depends on the value of $k$: high $k$-values require higher $\lambda$ values to achieve better accuracy.
The first term, that is the ASCE, is constant, while the second one, that is the AZPE, only improves on it for large $\lambda$-values and worsens it for smaller ones, as one would expect for an asymptotic expansion. 
The third term, beyond the AZPE, also exhibits a non-negligible contribution in the small-$\lambda$ regime.

We will now offer a physical interpretation of the above terms and make some considerations about their properties in the case of more general systems than the quantum ring model.
In the (infinitely) strong interaction limit, a given system behaves very rigidly, since the position of the reference electron determines the positions of all the remaining electrons. Upon application of a perturbation, the response of the system is instantaneous, or adiabatic, while maintaining its rigidity, unless special symmetries are present. This behavior is likely to apply to a wider class of systems, both with a uniform (such as the quantum ring) and a non-uniform density. On the other hand it is unclear whether the frequency-independence of $f_2$ is equally general as we already move away from the strictly correlated electron limit by introducing zero-point vibrations: thus the adiabaticity of $f_2$ for general systems is still an open issue. 
Finally, as already noted before, the third term $f_3$ of the expansion will be non-adiabatic for general systems.

\section{Conclusions}
\label{secConclusions}

In this work, we have considered an exactly solvable model consisting of two interacting electrons on a quantum ring. We focused on the response properties and calculated the energy spectrum, the excitation amplitudes, the density response function, and 
the exchange-correlation kernel of  time-dependent density functional theory. In the limit of strong interaction, we developed an asymptotic expansion in powers of the square root of the interaction strength for the response function and kernel. 
For the kernel we found that its leading terms are local in time but non-local in space.
This already shows that the commonly used adiabatic local-density, or semi-local, approximations will fail for such strongly correlated systems, since they are local both in time and space.
We compared the expansion for the kernel to a similar one obtained from the adiabatic-SCE formalism~\cite{Lani2016} which has the spatial non-locality built in. The leading term of the exact expansion was found to coincide 
with the adiabatic-SCE kernel derived in Ref.~\cite{Lani2016}. After working out the next order term, the so-called zero-point energy contribution, we found that it also coincided with the exact next-to-leading term. For our model, the subsequent term in the expansion is still adiabatic, but we showed that in general systems this term will be non-adiabatic.

The agreement with our exact results puts the adiabatic-SCE and the adiabatic-ZPE approximations on firmer ground and gives confidence in employing the formalism of strictly correlated electrons in the adiabatic approximation for calculating response properties of strongly correlated systems. 

\begin{acknowledgments}
LC, DK and RvL acknowledge the Academy of Finland for support under project no. 267839. 
GL acknowledges the Universit{\'e} Pierre et Marie Curie through an ATER position, which granted 
her considerable flexibility and freedom to contribute to this work. 
\end{acknowledgments}

\appendix
\section*{Appendices:}

\section{Response function of a harmonic model system}\label{App:HarmonicExample}

To illustrate some properties of the response function in the large interaction limit for an inhomogeneous
system, let us analyze a model system of two harmonically confined electrons in three dimensions with a harmonic repulsion\cite{Moshinksy1970}.
The Hamiltonian of the system is given by
\begin{equation}
\hat{H} = - \frac{1}{2} ( \nabla_1^2 + \nabla_2^2) + \frac{1}{2} \omega_\lambda^2 (|\br_1|^2 + |\br_2|^2) - \frac{\lambda}{2} \left |\br_1-\br_2 \right |^2
\end{equation}
in which the harmonic frequency $\omega_\lambda$ is chosen in such a way that the density is independent of $\lambda$.
Using the coordinate transformation $\bs=(\br_1 + \br_2) /\sqrt{2}$ and $\br=(\br_1-\br_2)/\sqrt{2}$ the Hamiltonian can be written as that of
two independent harmonic oscillators
\begin{equation}
\hat{H} = -\frac{1}{2} (\nabla_\bs^2 + \nabla_\br^2) +   \frac{1}{2}  \omega_\lambda^2 |\bs|^2 +
\frac{1}{2}  \nu_\lambda^2 |\br|^2 
\end{equation}
where $\nu_\lambda^2=\omega_\lambda^2- 2 \lambda$. The eigenfunctions and eigenvalues of this Hamiltonian are well-known. The normalized eigenfunctions are given by
\begin{equation}
\Psi_{\bn \bmm} (\bs,\br) = (\omega_\lambda \nu_\lambda)^{3/4} \Phi_\bn (\sqrt{\omega_\lambda} \, \bs) \Phi_\bmm (\sqrt{\nu_\lambda} \,\br) 
\end{equation}
where we defined the triplet of non-negative integers $\bn=(n_1,n_2,n_3)$ and  $\bmm=(m_1,m_2,m_3)$ and the functions
\begin{equation}
\Phi_\bn (\mathbf{x}) = \mathcal{H}_\bn (\mathbf{x}) \frac{e^{-|\mathbf{x}|^2/2} }{ \pi^{3/4} }
\nonumber
\end{equation}
in which we denoted
\begin{equation}
\mathcal{H}_{\bn} (\mathbf{x}) =  \frac{H_{n_1} (x_1) H_{n_2} (x_2) H_{n_3} (x_3)}{\sqrt{2^{|\bn|} n_1!n_2!n_3!}} 
\end{equation}
where $|\bn|=n_1+n_2+n_3$ and $H_n (x)$ is the Hermite polynomial of order $n$. The energy eigenvalues are given by
\begin{equation}
E_{\bn \bmm} = \omega_\lambda \left (|\bn| + \frac{3}{2} \right) + \nu_\lambda  \left(|\bmm| + \frac{3}{2}\right).
\end{equation}
The ground-state wave function $\Psi_0 = \Psi_{\mathbf{0}\mathbf{0}}$ has the explicit form
\begin{equation}
\Psi_0 (\bs,\br) = \left[ \frac{\omega_\lambda \nu_\lambda}{\pi^2} \right]^{3/4} e^{- \omega_\lambda |\bs|^2/2 -  \nu_\lambda |\br|^2/2 }.
\end{equation}
From this function the density is readily obtained as
\begin{equation}
n (\mathbf{x}) = 2 \left[ \frac{\beta}{\pi} \right]^{3/2} e^{- \beta |\mathbf{x}|^2} ,
\end{equation}
where we defined
\begin{equation}
\beta = \frac{2 \omega_\lambda \nu_\lambda}{\omega_\lambda + \nu_\lambda}.
\end{equation}
If we insert $\nu_\lambda = \sqrt{\omega_\lambda^2- 2 \lambda}$ into this relation, we can determine the $\lambda$-dependence of $\omega_\lambda$ 
as $\beta$ is independent of $\lambda$. We find
\begin{align}
\omega_\lambda &= \sqrt{\frac{\lambda}{2}} \, \left(y + \frac{1}{y}\right) \\
\nu_\lambda &= \sqrt{\frac{\lambda}{2}} \, \left(y - \frac{1}{y}\right) 
\end{align}
where $y$ solves the quartic equation
\begin{equation}
y^4 -  \beta \sqrt{\frac{2}{\lambda}} \, y^3 -1 = 0.
\end{equation}
In the limit of large interaction we find that $y=1+\beta /(2 \sqrt{2\lambda})+ \ordo{\lambda^{-1}}$ such that
\begin{align}
\omega_\lambda &= \sqrt{2\lambda} +\ordo{1} \\
\nu_\lambda &= \beta /2+ \ordo{\lambda^{-1/2}}
\end{align}
We see that the harmonic frequency of the center-of-mass mode approaches infinity
whereas the one of the relative mode approaches a 
finite value. This has interesting consequences for the excitation spectrum.
For the excitation energies of the relative mode it implies that 
\begin{equation}
\lim_{\lambda \rightarrow \infty} (E_{\mathbf{0} \bmm} - E_{\mathbf{00}} ) = \frac{\beta}{2} |\bmm|
\end{equation}
while all other excitation energies diverge to infinite values at large interaction strength.
The latter correspond to excitations of the center-of-mass mode. In contrast to the
quantum ring, only the excitation energies of the relative mode remain finite in
the large interaction limit which is due to the very different nature of the two-body interaction.

Let us turn our attention to the density response function. In the response function only the singlet excitations contribute. This means that the spatial wave functions that we need to consider are 
symmetric in the interchange of the particle positions. For this to be true, the relative wave functions need
to be even and we have to require that $|\bmm|$ only attains even values. The response function therefore
has the form
\begin{align}
\chi (\br_1,\br_2,\omega)=\sum_{ \bn, \bmm }&
\left[\frac{ D_{\bn \bmm} (\br_1)  D^*_{\bn \bmm} (\br_2)  }{\omega-\Delta E_{\bn \bmm}+i\eta}\right.\nonumber\\
-&\left.  \frac{ D_{\bn \bmm} (\br_2)  D^*_{\bn \bmm} (\br_1)  }{\omega + \Delta E_{\bn \bmm}+i\eta}   \right]
\label{ResponseHarmonic}
\end{align}
where $\Delta E_{\bn \bmm}=E_{\bn \bmm} - E_{\mathbf{00}}$ are the excitation energies and we further 
put the restriction that we sum over all $\bmm$ such that $|\bmm|$ is even.
The excitation amplitudes corresponding to these
excitations are given by
\begin{align}
D_{\bn \bmm} ( \br_1) &= \langle \Psi_0 | \hat{n} (\br_1) | \Psi_{\bn \bmm} \rangle \nonumber \\
&= 2 \int d\br_2 \Psi^*_{\mathbf{0} \mathbf{0}} (\br_1,\br_2)  \Psi_{\bn \bmm} (\br_1,\br_2)
\end{align}
where we rewrote the eigenfunctions in terms of the original coordinates.
Let us now consider the large interaction limit $\lambda \rightarrow \infty$ of the response function.
Since only the excitation energies of the form $\Delta E_{\mathbf{0} \bmm}$ remain finite in this limit
we only need to consider the excitation amplitudes of the form
\begin{align}
D_{\mathbf{0} \bmm} (\br_1) &=  2 \left[ \frac{\omega_\lambda \nu_\lambda}{\pi^2} \right]^{3/2} 
\int d\br_2 \, \mathcal{H}_{\bmm} (\sqrt{ \nu_\lambda/ 2} (\br_1 - \br_2) ) \nonumber \\
& \times e^{- \frac{\omega_\lambda}{2} (\br_1+\br_2)^2 -  \frac{\nu_\lambda}{2} (\br_1-\br_2)^2}.
\end{align}
If we now use that
\begin{equation}
\lim_{\lambda \rightarrow \infty} \left[ \frac{\omega_\lambda}{2 \pi} \right]^{3/2}  e^{- \frac{\omega_\lambda}{2} (\br_1+\br_2)^2 } = \delta (\br_1 + \br_2)
\end{equation}
is a limit representation for the delta distribution
and the fact that $\nu_\lambda \rightarrow \beta/2$ in the large interaction limit we find that
\begin{equation}
\lim_{\lambda \rightarrow \infty} D_{\mathbf{0} \bmm} (\br_1)  = 2 \left(  \frac{\beta}{\pi} \right)^{3/2} \mathcal{H}_\bmm (\sqrt{\beta} \, \br_1)
e^{- \beta |\br_1|^2}.
\end{equation}
For the response function in the large interaction limit we therefore obtain
\begin{align}
\alpha (\br_1,\br_2,\omega) &= \lim_{\lambda \rightarrow \infty}  \chi (\br_1,\br_2, \omega) = \nonumber \\
&n(\br_1 ) n (\br_2) \sum_{\bmm} 
\mathcal{H}_\bmm (\sqrt{\beta} \,\br_1)  \mathcal{H}_\bmm (\sqrt{\beta} \,\br_2) \nonumber \\
& \times  \left[ \frac{1 }{\omega -
\frac{\beta}{2} |\bmm | +  i \eta} \right. 
-  \left. \frac{1 }{\omega +
\frac{\beta}{2} |\bmm | +  i \eta} \right]
\label{eq:alfa_harm}
\end{align}
where the sum runs over $\bmm$ values such that $|\bmm|$ is even. We see that in the large interaction limit it
is still possible to excite the relative modes of the system. 

Let us now have a look at the external potential
in the strong interaction limit
\begin{equation}
v_\lambda (\br) = \frac{1}{2} \omega_\lambda^2 |\br|^2  = \lambda |\br|^2 + O (\sqrt{\lambda})
\end{equation}
which implies $\nabla v_\lambda (\br) = 2 \lambda \br + O (\sqrt{\lambda}) $. Since the response function does not
vanish in the large interaction limit, Eq.(\ref{eq:alfa_func}) tells us that
\[
\mathbf{0} = \int d\br_2 \, \alpha (\br_1,\br_2;\omega=0) \, \br_2 
\]
must hold, where we took the static limit of the response function of Eq.(\ref{eq:alfa_harm}). Since the polynomial functions in 
Eq.(\ref{eq:alfa_harm}) are all even (since $|\bmm|$ is even), and $\br_2$ is an odd function, we find that this relation is
indeed satisfied.

\section{Sips' expansion of the Mathieu function}\label{App:Sips}

In this Appendix we describe the details of the expansion of the Mathieu functions for large $q$. 
A recent general discussion is given by Frenkel and Portugal \cite{Frenkel2001} who give an overview of various expansions
for the Mathieu functions for large and small values of $q$ in different regions of their domain and recursion 
formulas to determine the expansion coefficients. The expansion that we are interested in for this work is the large $q$ expansion for the
Mathieu cosine function $C_l (z;q)$ in the region enclosing the value $z=\pi/2$.
Such an expansion was derived originally by Sips \cite{Sips1949,Sips1959,Blanch1960,NIST}
who developed a systematic theory.
For the Mathieu cosine this expansion is of the form of Eq.(\ref{eq:sips_cosine}). We here give the explicit expressions for the coefficients $c_{2n,l} (q)$ in Eq.(\ref{eq:sips_cosine}), and for this it will be convenient to define new coefficients $g_{2n,l} (q)$
\begin{equation}
c_{2n,l} (q) = \mathcal{C}_l (q) \, g_{2n,l} (q)
\label{eq:gcoeffs}
\end{equation}
which only differ from the coefficients $c_{2n,l} (q)$ by a prefactor $\mathcal{C}_l (q)$. This is done to ensure
that $g_{0,l} (q) =1$, which is convenient for a recursive calculation of the remaining coefficients $g_{2n,l} (q)$
as is done in \cite{Blanch1960,Frenkel2001}. The prefactor $\mathcal{C}_l (q)$ is chosen such that the Mathieu cosine satisfies
the normalization of Eq.(\ref{eq:mathieu_norm}). It has the explicit asymptotic expansion \cite{Sips1949,Blanch1960,Frenkel2001}
\begin{align}
\mathcal{C}_{l}(q)&=\left(\frac{\pi \sqrt{q}}{2 (l!)^2}\right)^{1/4}\left(1+\frac{2l+1}{8 \sqrt{q}} \right. \nonumber\\ 
&+ \left. \frac{l^4+2l^3+263 l^2+262 l+108}{2048 q}+  .. \right)^{-1/2} .\label{norm1} 
\end{align}
In terms of the new coefficients $g_{2n,l} (q)$ the Sips expansion of Eq.(\ref{eq:sips_cosine}) becomes \cite{Blanch1960}
\begin{equation}
C_l (z;q) = \mathcal{C}_l (q) \sum_{n=-\infty}^{\infty } g_{2n,l} (q) \, \mathcal{D}_{2n+l} (u)
\label{eq:sips_exp2}
\end{equation}
where the  functions $\mathcal{D}_l(u)$ are defined in Eq.(\ref{eq:Dfunc}) and $u =\sqrt{2} \, q^{1/4} \cos z$. 
Note that the expansion here differs in the choice of argument $u$ compared to that of Ref.~\cite{NIST} by a factor of $\sqrt{2}$ as we preferred to use
the physicist's convention for the Hermite polynomials appearing in the harmonic oscillator functions $\mathcal{D}_l$.
We prefer here to avoid a general discussion on the determination of $g_{2n,l} (q)$ and refer the interested reader to Refs.~\cite{Blanch1960,Frenkel2001} for details. Instead we give the explicit forms of the coefficients $g_{2n,l} (q)$ up to order $1/q$ which is sufficient for this work:
\begin{align}
g_{-8,l} (q) &= \frac{1}{2^{13} q} 8! \binom{l}{8}   \nonumber \\
g_{-6,l} (q) & = -\frac{1}{2^{10} q} 6! \binom{l}{6} \nonumber \\
g_{-4,l} (q) &=   \left[ \frac{1}{2^6 \sqrt{q}} + \frac{ (l-1)}{2^{8} q} \right] 4! \binom{l}{4}  \nonumber \\
g_{-2,l} (q) &= -\left[   \frac{1}{2^4 \sqrt{q}} + \frac{ (l^2 + 27l -10)}{2^{10} q} \right] l(l-1) \nonumber \\
g_{0,l} (q) &= 1 \nonumber \\
g_{2,l} (q) &= \left[  -\frac{1}{2^4 \sqrt{q}} + \frac{ (l^2-25 l -36)}{2^{10} q}    \right] \nonumber \\
g_{4,l} (q) &= \left[  -\frac{1}{2^6 \sqrt{q}} - \frac{  (l+2) }{2^{8} q}    \right] \nonumber \\
g_{6,l} (q) &= \frac{1}{2^{10} q} \nonumber \\
g_{8,l} (q) &= \frac{1}{2^{13} q} \nonumber 
\end{align}
where we define $\binom{n}{m} = 0$ if $n < m$.
From the formulas given in Ref.~\cite{Frenkel2001} it is readily seen that to know the Sips expansion to order $q^{-k/2}$ we need
to know the coefficients $g_{2n,l}(q)$ for $n \in \{-2k,\ldots,2k \}$. For example, the knowledge of the next order $q^{-3/2}$ would need the
knowledge of the coefficients from $g_{-12,l} (q)$ up to $g_{12,l} (q)$, see for example \cite{Blanch1960}.

\begin{figure}[ht]
  \centering
   \includegraphics[width=0.48\textwidth]{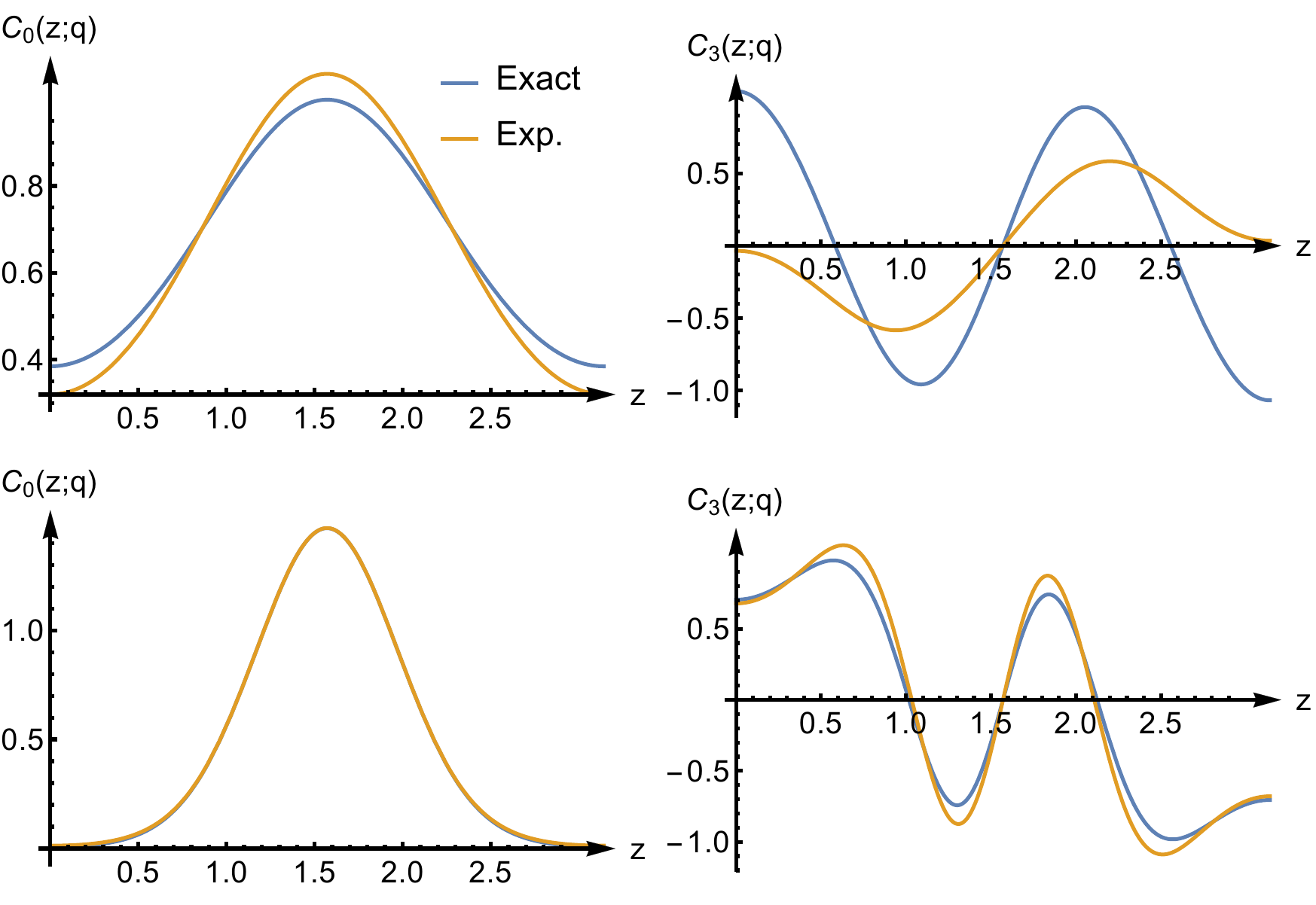}
  \caption{Comparison of the exact Mathieu functions to the expansion Eq. (\ref{eq:sips_exp2}) for the large $q$ limit in which the expansion coefficients
  $g_{2n,l}(q)$ that contain terms up to order $1/q$ were used. The upper (lower) panels have $q=1 \ (q=10)$, while the left (right) panels have $l=0 \ (l=3)$. 
  The accuracy of the expansion converges to the exact case quicker for lower values of $l$. 
  \label{ExactMathieuvsExpansion} }
\end{figure}

To show the accuracy of the Sips expansion we display in Fig.\ref{ExactMathieuvsExpansion} the Mathieu cosine $C_l (q;z)$ for
$l=0$ and $l=3$ for $q=1$ and $q=10$ and compare them to the Sips expansion using the coefficients $g_{2n,l} (q)$ given explicitly above.
We see that already at $q=10$ the Sips expansion performs very well. In general we find that for higher values of $l$ more terms in the
expansion need to be taken into account for high accuracy. 

It remains to give the asymptotic values of the Mathieu characteristic value.
The characteristic values $a_l^+(q)$ and $a_l^-(q)$ have the same asymptotic expansion of
the form~\cite{Frenkel2001}
\begin{align}
a_l^{\pm} (q)&=-2 q+2 (2l+1) \sqrt{q}-\frac{1}{4}\left(2 l^2 + 2l+1\right) \nonumber \\
&- \frac{(2l+1)}{128 \sqrt{q}}\left((2l+1)^2+3\right) + \ordo{q^{-1}} \label{eq:excitationEnergies}
\end{align}
while the difference $a_l^+ (q)-a_l^- (q)$ is exponentially small in the large $q$ limit \cite{NIST}. 

\section{Expansion of the excitation amplitudes} \label{App:expansion}

\label{AppDq}

Here we further outline some general features of the expansion of the excitation amplitudes $D_{kl} (q)$.
If we rewrite the expansion of Eq.(\ref{eq:Dkl_exp}) using the coefficients of Eq.(\ref{eq:gcoeffs}) we have the expression
\begin{align}
D_{kl} (q) &= \frac{2}{\pi} \mathcal{C}_0 (q) \, \mathcal{C}_l (q)  \! \! \! \! \! \! \! \!
\sum_{n_1,n_2 = -\infty}^\infty  \! \! \! \! \!  g_{2n_1,0} (q) \, g_{2n_2,l} (q) \mathcal{J}_{kl}^{n_1 n_2} (q)
\label{eq:Dexp2}
\end{align}
For the products of the pre-factors we can write
\begin{equation}
 \mathcal{C}_0 (q) \mathcal{C}_l (q) = q^{1/4} \, F_l (q) \label{C_product}
\end{equation}
where $F_l (q)$ has an expansion in powers of $q^{-1/2}$, i.e.
\begin{equation}
F_l (q) = \frac{1}{l!} \sqrt{\frac{\pi}{2 } } \left[ 1 - \frac{l+1}{8 \sqrt{q}} + \ordo{q^{-1}} \right]
\end{equation}
and higher powers can be calculated from expression (\ref{norm1}).
With these definitions and Eq.(\ref{eq:Jcoeff}) we can rewrite the expansion of Eq.(\ref{eq:Dexp2}) as
\begin{align}
& D_{kl} (q) = \frac{2}{\pi} F_l (q) \nonumber \\
& \times \sum_{n_1,n_2 = -\infty}^\infty g_{2n_1,0} (q) g_{2n_2,l} (q) \sum_{r=0}^\infty   \frac{a_r (k)  I_{n_1 n_2, r}^l
}{(\sqrt{2})^{r+1} q^{r/4} }
\end{align}
where the factor $q^{1/4}$ from Eq.(\ref{C_product}) has been absorbed in the last sum.
The function $F_l (q)$ and the coefficients $g_{2n,l} (q)$ have an expansion in powers of $q^{-1/2}$. Now since the coefficients 
$I_{n_1 n_2,r}^l$ vanish unless $k$ and $l$ are both even or both odd, we conclude that $D_{kl} (q)$ has an expansion only in odd powers
of $q^{-1/4}$ if $l$ is odd and only in even powers of $q^{-1/4}$ otherwise. The explicit expression for $I_{n_1 n_2,r}^l$ is given by
\begin{widetext}
\begin{align}
I_{n_1 n_2,r}^l &=  \int_{-\infty}^\infty du \, u^r \,  \mathcal{D}_{2n_1} (u) \mathcal{D}_{2n_2+l} (u) 
= \left\{ \begin{array}{cc}  0 & \mbox{If $r+l$ odd} \\ 
\frac{r!}{2^r} \sqrt{\pi} \, 2^{n_1+n_2 +\frac{l}{2}}
\sum_{p= \max{(0,-s)}}^{\min (2n_1, 2n_2+l)} \binom{2n_1}{p} \binom{2n_2+l}{p} \frac{p!}{2^p (s+p)! }  & \mbox{otherwise}    
\end{array}  \right.  \nonumber
\end{align}
\end{widetext}
where we defined $s=r/2 - n_1 - n_2 - l/2$ \cite{wolfram}.
Finally the coefficients $a_r (k)$ can be obtained from a Taylor expansion of the function $f_k (x)$ defined in Eq. (\ref{eq:fkx}). Taken all these terms
together we find the following expansion of the excitation amplitude:
\begin{widetext}
\begin{align}
D_{kl}(q)= 
&(-i)^k \left[ 1 - \frac{k^2}{8 \, \sqrt{q}} + \frac{k^2 (k^2-4)}{128 \,q}  - \frac{k^2 (k^4 - 14 k^2 + 46)}{3072 \, q^{3/2}}  + \ordo{q^{-2}}\right] \, \delta_{l0}  \nonumber \\
+ 
&(-i)^{k+1} \left[ -\frac{k}{2 q^{1/4}} + \frac{k (k^2-1)}{16 \, q^{3/4}}  - \frac{k (2 k^4 - 16 k^2 +13)}{512 \, q^{5/4}} + \ordo{q^{-7/4}} \right] \,
\delta_{l1}  \nonumber \\
+ 
&(-i)^k \left[ - \frac{k^2}{4\sqrt{2} \sqrt{q}} +  \frac{k^2 (2 k^2-5)}{64 \sqrt{2} \, q}   + \ordo{q^{-3/2}}\right]  \, \delta_{l2} + (-i)^{k+1} \left[  \frac{k (2k^2 +1)}{16\sqrt{6} \, q^{3/4} }  
+ \ordo{q^{-5/4}} \right]\, \delta_{l3}   \label{eq:Dkl}
\end{align}
\end{widetext}
The excitation amplitude indeed has an expansion in odd powers of $q^{-1/4}$ for odd $l$ and in even powers of $q^{-1/4}$ for even $l$, as we demonstrated above. 
This implies that $|D_{kl} (q)|^2$ has an expansion in powers of $q^{-1/2}$. The expansion derived here has been checked numerically to ensure the correctness
of the derivations.
As a further check on the result we see that the excitation amplitude satisfies $D_{kl} (q)=(-1)^{k+l} D_{kl}^{*} (q)$ as well as $D_{0l} (q) = \delta_{l0}$.

The order for $D_{kl}(q)$ given in \Eq{eq:Dkl} is enough to yield the following asymptotic expansion for the absolute value squared of the excitation amplitudes to order $q^{-3/2}$
\begin{widetext}
\begin{align}
|D_{kl}(q) |^2= & \left[ 1 - \frac{k^2}{4 \, \sqrt{q}} + \frac{k^2 (k^2-2)}{32 \,q} - \frac{k^2 (2 k^4 - 13 k^2+ 23)}{768 \, q^{3/2}}  \right] \, \delta_{l0} +  \left[ \frac{k^2}{4 \sqrt{q}} + \frac{k^2 (1-k^2)}{16 \, q}
+  \frac{k^2( 4 k^4 - 20 k^2 +15)}{512 \, q^{3/2}}  \right] \, \delta_{l1} \nonumber \\ 
+  &\left[ \frac{k^4}{32 \,q}- \frac{k^4 (2k^2- 5)}{256 \, q^{3/2} }  \right] \, \delta_{l2} +  \left[  \frac{k^2 (4 k^4 + 4 k^2 +1)}{1536 \, q^{3/2}} \right] \, \delta_{l3} + 
\ordo{q^{-2}}.
\label{eq:D2}
\end{align}
\end{widetext}

\begin{figure}[ht]
  \centering
   \includegraphics[width=0.48\textwidth]{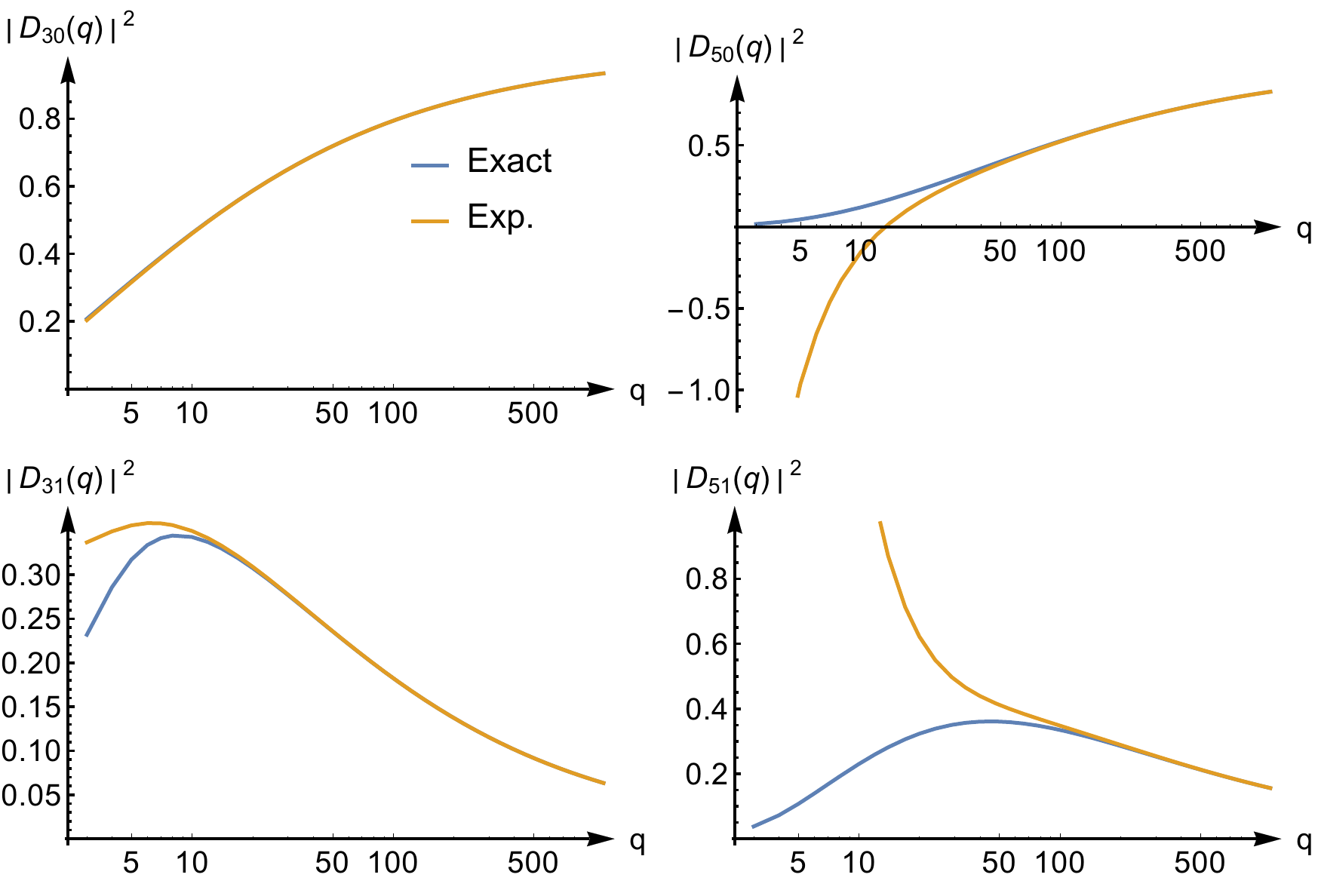}
  \caption{Square $|D_{kl}(q)|^2$ of the excitation amplitudes for $l=0$ (top) and $l=1$ (bottom) for $k=3$ (left) and $k=5$ (right) as function of $q$. We compare the square of Eq.(\ref{dintegral1}) (denoted by Exact) to the expansion of Eq.(\ref{eq:D2})
  (denoted by Exp.). 
  \label{ComparisonDkl} }
\end{figure}

We compare the expansion of $|D_{kl}(q) |^2$ to the exact values obtained by numerically integrating \Eq{dintegral1} in Fig.\ref{ComparisonDkl}. 
We see from this figure that higher values of $k$ require higher values of $q$ to make the expansion accurate. This was to be expected since larger
$k$-values imply a more oscillatory integrand in Eq.(\ref{dintegral1}) while a larger value of $q$ makes the Mathieu functions more localized and thereby
making the expansion of Eq.(\ref{eq:fkx}) used in the integrand more accurate. 

We finally note that the frequency-sum rule, \Eq{eq:f_sumrule}, can be used to check the validity of some of the terms of \Eq{eq:D2}. If we insert the explicit form of Eq.(\ref{eq:chi_k}) into the 
one-dimensional equivalent of Eq.(\ref{eq:f_sumrule}) for the
frequency sum rule we can derive that
\begin{equation}
\sum_{l} \, ( k^2 + [ a_l^+ (q) - a_0^+ (q) ] ) |D_{kl} (q)|^2 = k^2
\end{equation}
where the sum runs over even $l$ if $k$ is even and over odd $l$ when $k$ is odd. The right hand side is independent of the interaction strength $q$, thus in the sum on the left hand side the $q$-dependence in the excitation amplitudes has to be 
compensated by the $q$-dependence of the Mathieu characteristic values to give a result independent of the interaction strength.

%

\end{document}